\documentclass[fleqn,usenatbib]{mnras}

\usepackage{newtxtext,newtxmath}

\usepackage[T1]{fontenc}

\DeclareRobustCommand{\VAN}[3]{#2}
\let\VANthebibliography\thebibliography
\def\thebibliography{\DeclareRobustCommand{\VAN}[3]{##3}\VANthebibliography}

\usepackage{graphicx}	
\usepackage{amsmath}	
\usepackage{natbib}
\usepackage{xcolor}

\title[{[C II]} as a tracer of stellar feedback]{Stellar feedback SPICEs up [C II] emission in the first galaxies}

\author[Bhagwat et al.]{
Aniket Bhagwat,$^{1}$\thanks{E-mail: abhagwat@mpa-garching.mpg.de}, 
Tiago Costa$^{1,2}$,
Benedetta Ciardi$^{1}$ and 
Fabrizio Arrigoni Battaia$^{1}$
\\
$^1$Max Planck Institut f\"{u}r Astrophysik, Karl Schwarzschild Stra{\ss}e 1, D-85741 Garching, Germany\\
$^2$School of Mathematics, Statistics and Physics, Newcastle University, UK
}
\date{Accepted XXX. Received YYY; in original form ZZZ}

\pubyear{2015}

\begin{document}
\label{firstpage}
\pagerange{\pageref{firstpage}--\pageref{lastpage}}
\maketitle

\begin{abstract}
The bright [C II] 158 $\mu$m line is a common tracer of star-forming gas and feedback-driven outflows in high-redshift galaxies. We use the {\tt SPICE} simulations to quantify how variations in the behavior of stellar feedback, including bursty and continuous energy injections, shape properties of $z \gtrsim 5$ galaxies as traced by [C II] emission. In all models, we find a tight correlation between [C II] luminosity ($L_{\rm [CII]}$) and star formation rate. However, bursty feedback results in systematically lower $L_{\rm [CII]}$ at fixed star formation rate, and larger intrinsic scatter. [C II] emission is systematically more extended than the rest-frame UV continuum by factors of $\approx 2{-}4$, in agreement with ALMA observations at $z>5$. Although outflowing motion is ubiquitous and mass outflow rates scale with $L_{\rm [CII]}$, reaching characteristic values of $\sim 10\,M_\odot\,{\rm yr^{-1}}$ the net mass flux is inflow dominated for [C II] bright galaxies, even if feedback is bursty. Radial velocity distributions show that [C II] emission is dominated by low-velocity ($|v_{\rm rad}| < 200 \rm \ km/s$), dynamically cold gas, while the fast moving gas does not emit [C II]. As a result, [C II]-based velocity estimates do not recover the full multiphase structure of galactic winds: they overestimate the bulk velocity of the cold gas while underestimating the characteristic velocity of the fast outflow by factors of $\sim2$--5., suggesting that [C II] is biased tracer of outflowing winds. Consequently, mass-loading factors and wind energetics inferred from [C II] are systematically biased low. The fastest outflowing gas is not [C II] bright, and the largest [C II] line widths are instead set by the gravitational potential. While the spectral and spatial distribution of [C II] alone does not clearly distinguish between feedback models, gas kinematics provides a strong diagnostic. Smooth feedback promotes earlier disk settling, yielding higher $V/\sigma$ and disk fractions ($\sim48\%$ for $M_* > 10^8 \rm M_\odot$ at $z=5$) compared to bursty feedback ($\sim28\%$) demonstrating an observable imprint of feedback on galaxy kinematics. Overall, while [C II] robustly traces star formation, it does not reliably trace feedback-driven outflows, underscoring the need for multiwavelength diagnostics (e.g. ALMA and JWST) to constrain the full multiphase structure of high-redshift galaxies.

\end{abstract}

\begin{keywords}
Galaxies : ISM -- galaxies : reionisation -- galaxies : infrared emission line -- methods : numerical
\end{keywords}


\section{Introduction}
The $158\mu \rm m$ fine-structure line arising from the $^2P_{3/2} \rightarrow ^2P_{1/2}$ transition of singly ionised carbon ([C II] henceforth) has emerged as a key probe to study the gas in and around high-redshift galaxies. [C II] is a dominant coolant in the interstellar medium (ISM) \citep{Malhotra1997,Luhman1998} and accounts for $\sim$$1\%$ of the total far-infrared (FIR) luminosity of galaxies \citep{Russell1980CIIdetection,Spitzer1978,Crawford1985CII,Stacey1991CII}. The versatility of this line lies in its ability to trace a variety of phases of the interstellar medium (ISM), including photo-dissociation regions (PDRs), diffuse ionised gas and the cold neutral medium \citep{2019Anderson,Tanantino2021}. Each of these phases can excite [C II] emission due to its low ionisation potential of $11.2 \rm eV$ \citep{Madden1993,Pineda2013,Croxall_2017,Tanantino2021}. [C II] may therefore act as a diagnostic of the structure and kinematics of the multi-phase gas within the halos of star-forming galaxies.

At $z\geq 1$, [C II] redshifts into sub-millimeter wavelengths allowing state-of-the-art observing facilities such as the Atacama Large Millimeter/sub-millimeter Array (ALMA) \citep{ALMA} and the Northern Extended Millimeter Array (NOEMA) to observe [C II] emitters up to $z\sim11$ \citep{Aravena_2016,FudamotoGNz112024}. In the last few years, multiple large ALMA programs such as {\tt ALPINE} \citep{LeFevre2020,Bethermin2020Alpine,Faisst2020Alpine}, {\tt REBELS}  \citep{Bouwens2022Rebels} and {\tt CRISTAL} \citep{Solimano2024,mitsuhashi2023almacristal} have observed hundreds of galaxies in the heart of the epoch of reionisation, between $4<z<7$. These programs aim to understand the physical origin of [C II] emission, its connection to star-formation, metal enrichment in the first galaxies, mass assembly via mergers, incidence of cold outflows, gas kinematics and dynamics in the galaxies that drive cosmic reionisation. Additionally, these programs allow for verification of empirical scaling relations between [C II] and physical properties of galaxies observed in the local universe at high redshifts \citep[e.g.][]{peng2025, Decarli2025}.  

A strong correlation between [C II] luminosity ($L_{\rm CII}$) and the star-formation rate (SFR) has been shown by observations of galaxies in the local universe \citep{Boselli_2002,DeLooze2011SFRCII,HerreraCamusCIISFR2015,Hodge_2020}, suggesting that $L_{\rm CII}$ is a strong tracer of SFR. At $z>5$, however, a deficit in the $L_{\rm CII}$-SFR relation or a non-detection of [C II] in star-forming galaxies has raised questions on the use of [C II] as a tracer of star-forming gas,  with some groups arguing toward a deficit in luminosity \citep{Ouchi_2013CIIdeficit,Ota2014CIIdeficit,Maiolino2015CIIdeficit,Knudsen2016CIIdeficit,Pentericci2016CIIdeficit,Bradač_2017CIIdeficit,Laporte2017CIIdeficit,2021Jolly,Romano2022CIIdeficit}, while others claim a lack of a deficit and consistency with low-redshift scaling relations \citep{2017Matthee,Carniani_2020,Schaerer_2020,Fujimoto_2019}. Numerical studies also demonstrate similar disagreements, where some authors show the presence of a luminosity deficit due to thermal saturation \citep{2016Munoz,2017Narayanan}, enhanced interstellar radiation field \citep[ISRF;][]{2018LagacheL,Bisbas_2022,2024Ebagezio}, short depletion times or low metallicities \citep{2024Liang} and bursty star-formation \citep{Ferrara2019CII,Pallottini_2019,Katz_2022CII,Casavecchia_2024}, while other studies do not find any deficits in luminosity \citep{Olsen_2017,Leung2020,Richings_2022,2024SchimekS}.

The [C II] line, additionally, is a cornerstone in the study of star-formation-driven physical processes in the ISM, such as mechanical feedback \citep{2018Gallerani,2020Ginolfi}, chemical enrichment \citep{2020GinolfiCGM,Di_Cesare_2024} and kinematics of galaxies \citep{2017JonesCII,Jones_2021,telikova2024,rowland2024rebels25} at $z \gtrsim 4$. Cold inflows and outflows imprint their signatures on the [C II] line profiles via spectral broadening and presence of line wings \citep{Janssen_2016,Jones_2017Fil}. Therefore, individual [C II] line profiles can be exploited to constrain incidence of outflows, outflow velocities, and to infer mass flow rates  \citep{Herrera_Camus_2021,Herrera_Camus_2022}. This is only possible given the high signal-to-noise (S/N) of the data, and various authors resort to stacking analysis to boost signal and find evidence of outflows using the broad component of the stacked emission \citep{2018Gallerani,2019Sugahara,2020Ginolfi,birkin2025alma}. Additionally, multiple studies have shown that at $z>4$ [C II] emission is typically more extended than the rest-frame UV continuum observed with HST—by a factor of more than two—both in individual galaxies \citep[e.g.][]{Carniani_2018,Fujimoto_2020,2024IkedaI} and in stacking analysis \citep[e.g.][]{Fudamoto_2022,fudamoto2023ext}. The presence of extended two-component [C II] emission beyond the rest-frame UV indicates metal-enriched gas reaching CGM scales (out to $\sim 10 \rm kpc$) due to outflows in the first galaxies. 

The reliability of [C II] as a tracer of outflows, however, remains debated \citep{Spilker_2020,birkin2025alma}, as the origin of both the broad component and a two-component halo profile can also be attributed to other physical scenarios, such as satellite galaxies and mergers \citep{2020Jones,Kade_2023,2024IkedaI,Solimano2024,2024Posses} and cold streams and inflows \citep{Jones_2017Fil} or dynamical origins \citep{2015Kimball}. 
Observationally, these physical scenarios have been proven difficult to disentangle due to the presence of [C II] to different phases of gas, and to the typically poor spatial resolution ($1''$ $\approx 6.5$ kpc at $z=5$), hindering accurate morphological measurements. Therefore, a systematic study of [C II] line kinematics in different outflow scenarios is key toward understanding the efficacy of [C II] as a tracer of gas dynamics. 

[C II] datacubes and resulting moment maps play a crucial role in understanding the kinematics of the first galaxies. Using 3D tilted ring fitting codes such as {\tt BAROLO3D} \citep{Teodoro_2015} or 2D kinematic fitting codes such as {\tt GALPAK3D} \citep{Bouche_2015} and  {\tt DYSMAL} \citep{Cresci_2009} has become common practice to get an estimate of the rotational velocities and velocity dispersion of [C II] bright galaxies. Studies reveal a contrasting picture of the morpho-kinematic properties of the first galaxies with some studies suggesting a diversity of kinematics \citep{Jones_2021,telikova2024}, while others suggesting ubiquitous incidence of dynamically cold disk galaxies \citep{Rizzo:2020ijt,2021Fraternali,Lelli_2021,rowland2024rebels25,fujimoto2025p,scholtz2025}. The picture is however complicated by follow-up high resolution observations of previously observed "cold disks" which demonstrate complex morphology instead, as in the case of {\tt HZ4} \citep{Herrera_Camus_2021,Parlanti_2025} and {\tt CRISTAL-05} \citep{Fujimoto_2020, 2024Posses}. Resolution studies \citep{Rizzo_2022} and code comparision projects \citep{lee2024codes} have further raised the need to characterise systematics of [C II] kinematics with simulation models. 

A range of analytic \citep{2016Munoz,Ferrara2019CII} and semi-analytic \citep{Abel2007,Olsen_2017,2018LagacheL,Popping_2019,Pizzati_2022outflow,2024Liang} models of gas clouds and PDRs have been instrumental in exploring the physical processes governing [C II] luminosity. These models enable broad parameter space studies but are limited in their ability to resolve sub-galactic structures or predict statistical galaxy-scale observables. More recently, hydrodynamical simulations have become the standard approach \citep{Pallottini_2017,Katz_2017_alma,Lupi_2020,Katz_2019_FIR,Richings_2022,Bisbas_2022,Casavecchia_2024,2024SchimekS,2024Khatri,Mu_oz_Elgueta_2024}, despite their high computational cost. These simulations capture the complex, multiphase nature of galactic halos, enabling predictions of [C II] observables across ISM to CGM scales down to $\sim10 \ \rm pc$. While much of this work has focused on the origin, morphology, and kinematics of [C II] emission, a key open question remains: how do different modes of stellar feedback influence [C II] observables, and can [C II] serve as a diagnostic to constrain feedback models? 

In this paper, we exploit the feedback variations in the radiation-hydrodynamical suite of simulations called {\tt SPICE} \citep{bhagwat2024}, to study the connection between stellar feedback and [C II] observables of high-redshift galaxies. The paper is structured as follows: in section~\ref{sec:simulationsCII} we briefly describe the {\tt SPICE} simulations employed in this work, in sections~\ref{sec:CIImodel} and~\ref{sec:SyntheticdataCII} we introduce a model to predict [C II] luminosities from galaxies and model synthetic datacubes for ALMA observations and extract observables respectively, in section~\ref{sec:resultsCII} we present our results pertaining to feedback dependant [C II] properties of galaxies, in section~\ref{sec:discussion} we discuss the implications of our findings and connect them to existing literature, and finally we give our conclusions in section~\ref{sec:conclusionsCII}. 

\section{Methodology}
\label{sec:MethodsCII}
Here we describe the cosmological radiation-hydrodynamical simulations employed in this work, and introduce a post-processing model to estimate [C II] emission from simulated galaxies. 
\subsection{{\tt SPICE} Simulations}
\label{sec:simulationsCII}

We employ the {\tt SPICE} suite of radiation-hydrodynamical (RHD) simulations to study the impact of different models of supernova feedback on [C II] properties of galaxies. We briefly introduce the simulations below, and refer the readers to \citet[][hereafter B24]{bhagwat2024} for details. {\tt SPICE} is a set of RHD simulations performed using {\tt RAMSES-RT} \citep{rosdahl2013ramses,rosdahl2015scheme}, 
which is a coupled RHD extension of the adaptive mesh refinement (AMR) code {\tt RAMSES} \citep{teyssier2002cosmological}. {\tt RAMSES-RT} solves the coupled gas hydrodynamics, self-gravity and radiative transfer of stellar radiation along with the N-body dynamics of dark matter and stars. 

Initial conditions are generated using {\tt MUSIC2-monofonIC} \citep{Hahn_2020,Michaux_2020} at $z=30$ with the 2PPT/2LPT approach in a cosmological box of length 10 cMpc/$h$. A $\rm \Lambda CDM$ cosmological model is adopted, with parameters $\Omega_{\Lambda} = 0.6901$, $\Omega_{\rm m} = 0.3099$,  $\Omega_{\rm b} = 0.0489$, $H_0 = 67.74$ \ \rm km/s/Mpc, $n_{\rm s} = 0.9682$, $\sigma_8 = 0.8159$ \citep{Planck2016}, where the symbols have their usual meaning. Primordial gas mass fractions are $X=0.7545$ for Hydrogen (H) and $Y=0.2455$ for Helium (He), along with a metallicity floor of $Z = 3.2 \times 10^{-4}$ Z$_\odot$. {\tt SPICE} models the non-equilibrium metal line cooling at $T>10^4 \rm K$ using {\tt CLOUDY} tables, while at $T\leq10^4 \rm K$ it adopts fine structure rates from \citet{rosen1995global}. {\tt SPICE} also follows the non-equilibrium ionisation states of H and He ({\textsc{HII} , \textsc{H}e\textsc{II} , \textsc{H}e\textsc{III}}), which are advected as passive scalars and fully coupled to the local ionising radiation field \citep[see][]{rosdahl2013ramses}. 

The local resolution of the AMR grid is increased if one of the two following criteria are met: (i) $M_{\rm dm} + \frac{\Omega_{\rm m}}{\Omega{\rm b}} \ M_{\rm b} > 8\times m_{\rm dm}$ (quasi-Lagrangian refinement criteria), where $M_{\rm dm}$ and $M_{\rm b}$ are the total dark matter and baryonic (i.e. gas + stars) mass within the cell, and $m_{\rm dm}\approx6\times10^5 \ \rm M_\odot$  is the mean dark matter particle mass, or (ii) the cell width is larger than $\frac{1}{8}$ of the local Jeans length. This allows for a maximum resolution of $\approx 32 \rm pc$ at $z=5$, with a minimum resolution of $\approx4.8 \rm kpc$ at the coarsest refinement levels. 

{\tt SPICE} employs the multi-freefall star formation model implemented by \citet{kretschmer2020forming}, which includes a prescription to track sub-grid turbulence in each computational cell, yielding a variable star formation efficiency (SFE) that depends on the physical state of the gas in a cell as described by its virial parameter and turbulent mach number. Supernova (SN) feedback is included in the form of a mechanical feedback scheme introduced by \cite{kimm2014escape} (implemented as in \citealt{kimm2015towards}). The model injects momentum in the surrounding cells based on the phase of the SN remnant that is resolved. Depending on whether the Str{\"o}mgren radius is resolved or not, we additionally include a momentum correction to account for pre-processing of gas due to photo-ionisation \citep{geen2015detailed}. We assume a \citet{chabrier2003galactic} initial mass function (IMF), which implies that a fraction $\eta_{\rm sn} = 0.31$ of the initial stellar mass is recycled back into the ISM, and 5\% of this is ejected as metals heavier than He. The choice of IMF also results in a fixed SN rate of 0.016 SN $\rm M^{-1}_\odot$.  

While keeping the global implementation of the SN feedback constant, we systematically vary the timing and energies of SN events, as briefly described below (see Table 2 in \citealt{bhagwat2024}):
\begin{enumerate}
\item {\tt \textbf{bursty-sn}}:  All SN from a stellar particle are injected with a fixed energy of $E_{\rm SNII}=2\times10^{51}$ ergs in a single event when the particle reaches an age of 10 Myr. 
\item {\tt \textbf{smooth-sn}}:  SN from a stellar particle are injected with a fixed energy of $E_{\rm SNII}=2\times10^{51}$ ergs, but the SN events occur with a delay time distribution, i.e. between 3 and 40 Myr after the birth of a particle. 
\item {\tt \textbf{hyper-sn}}: For each stellar particle a fraction $f_{\rm HN}$ of SN explodes as hypernovae, injecting an energy of $10^{52}$ ergs. Following \citet{grimmett2020chemical}, we assume a metallicity dependent fraction $f_{\rm HN} = {\rm max}(0.5 \times \ {\rm exp}(-Z_\ast/0.001), 0.01)$, with $Z_\ast$ stellar metallicity. The remainder of the SN events inject energies between $10^{50} - 2\times10^{51}$ ergs \citep{sukhbold2016core}. The SN time delay distribution remains such that explosions occur between 3 and 40 Myr. 
\end{enumerate}

{\tt SPICE} also tracks the on-the-fly transport of radiation from stars in five photon frequency groups, i.e. infrared ($0.1-1 \ \rm eV$), optical (OP; $1-13.6 \ \rm eV$) and three bands of ionising ultraviolet radiation (UVI with $13.6-24.59$ eV, UVII with $4.59-54.42$ eV, UVIII with $54.42-\infty$ eV). The luminosity of stellar particles is evaluated depending on their age, metallicity and mass using SED models of BPASSv2.2.1 \citep{bpassv211,bpassv221}. Finally, {\tt SPICE} includes radiative feedback via photo-ionisation, photo-heating and direct radiation pressure from UV photons, as well as radiation pressure on dust from infrared and optical photons. The dust number density is modelled as a function of the local metallicity and H I number density, $n_{\rm d} \equiv (Z/Z_\odot)n_{\rm HI}$. Each cell is assigned dust absorption and scattering opacities (see Table 3 in \citealt{bhagwat2024}), allowing for re-processing of photons into the infrared group and for interaction with gas via multi-scattered radiation pressure \citep{rosdahl2015scheme}.
\subsection{Modelling [C II] emission}
\label{sec:CIImodel}
As {\tt SPICE} does not actively track the non-equilibrium chemistry and radiation coupling for the C II ionisation state, we post-process the snapshots to assign an emissivity to each computational cell, and calculate the luminosity emanating from it. As a carbon atom can exist in multiple ionisation states depending on the strength of the radiation field\footnote{The ionisation potentials of C I and C II are $11.2 \rm eV$ and $24.4 \rm eV$, respectively.} and the physical conditions of a cell, for each cell we write the equilibrium equation for a three-state carbon atom (i.e. carbon can exist in singly, doubly and triply ionised states) as: 
\begin{equation}
\begin{split}
    n_{\rm CI} \Gamma_{\rm CI} + n_{\rm CI} n_e \beta_{\rm CI} &= n_e n_{\rm CII} \alpha_{\rm CII}   \\
    n_{\rm CII} \Gamma_{\rm CII} + n_{\rm CII} n_e \beta_{\rm CII} &= n_e n_{\rm CIII} \alpha_{\rm CIII},
    \label{eq:detailedbal}
\end{split}
\end{equation}
where $n_{\rm CI}$, $n_{\rm CII}$ and $n_{\rm CIII}$ are the number densities of the three ionisation states, $\Gamma_{\rm CI}$ and  $\Gamma_{\rm CII}$ the photo-ionisation rates of C I and C II, $\beta_{\rm CI}$ and $\beta_{\rm CII}$ the collisional excitation rates \citep{2005Tielens}, $\alpha_{\rm CI}$ and $\alpha_{\rm CII}$ the recombination rates \citep{Badnell_2006}, and $n_e$ is the electron number density. We additionally require that $n_{\rm CI} + n_{\rm CII} + n_{\rm CIII} = 10^{-3.31} n Z\equiv n_{\rm C}$, where $10^{-3.31}$ is the fractional abundance of carbon in a cell \citep{2006Nomoto}, $n$ is the number density of gas, $Z$ is the metallicity and $n_{\rm C}$ is the number density of carbon. Therefore, one can derive $n_{\rm CII}$ as:
\begin{equation}
    n_{\rm CII} = \frac{ n_{\rm C}}{\left( 1 + \frac{\Gamma_{\rm CII} + n_{\rm C}\beta_{\rm CII}}{n_e \alpha_{\rm CIII}} + \frac{n_e \alpha_{\rm CII}}{\Gamma_{\rm CI} + n_e \beta_{\rm CI}}  \right)}.
\end{equation}
As {\tt SPICE} performs an on-the-fly radiation transport, the photo-ionisation rates for C I and C II can be extracted directly from the simulations, without any further assumption.
More specifically, as the OP frequency bin in {\tt SPICE} tracks radiation in the range $(1-13.6) \rm eV$, one can calculate the fraction of flux with $h \nu > 11.2 \rm eV$  by interpolating the stellar SEDs based on the current age, mass and metallicity of all the stars in a given cell. We then define a correction factor $f_{\rm CI} = N_{h\nu > 11.2} / N_{\rm OP}$, where $N_{h\nu > 11.2}$ and $N_{\rm OP}$ are the number of photons emitted by each star at $h \nu>11.2$~eV and in the OP band, respectively. Within a given cell, the local photo-ionisation rate of C I can then be evaluated as $\Gamma_{\rm CI} = \sigma_{\rm CI} \overline{f}_{\rm CI} F_{\rm OP}$, where $\sigma_{\rm CI}$ is the flux-weighted ionisation cross-section of C I \citep{1996Verner}, $\overline{f}_{\rm CI}$ is the luminosity weighted correction factor defined previously, and $F_{\rm OP}$ is the flux from the optical frequency bin obtained directly in {\tt SPICE}.
Finally, the photo-ionisation rate of C II can be evaluated as $\Gamma_{\rm CII} = \sigma_{\rm CII} F_{\rm UVII+UVIII}$, where $\sigma_{\rm CII}$ is the flux-weighted ionisation cross-section of C II \citep{1996Verner}, and $F_{\rm UVII+UVIII}$ is the flux from the UV II and UV III frequency bins obtained in {\tt SPICE}. In this case, no correction factor is needed as {\tt SPICE} already tracks exactly radiation with $h\nu > 24.4 \rm eV$.
\begin{figure*}
    \centering
    \includegraphics[width=\linewidth]{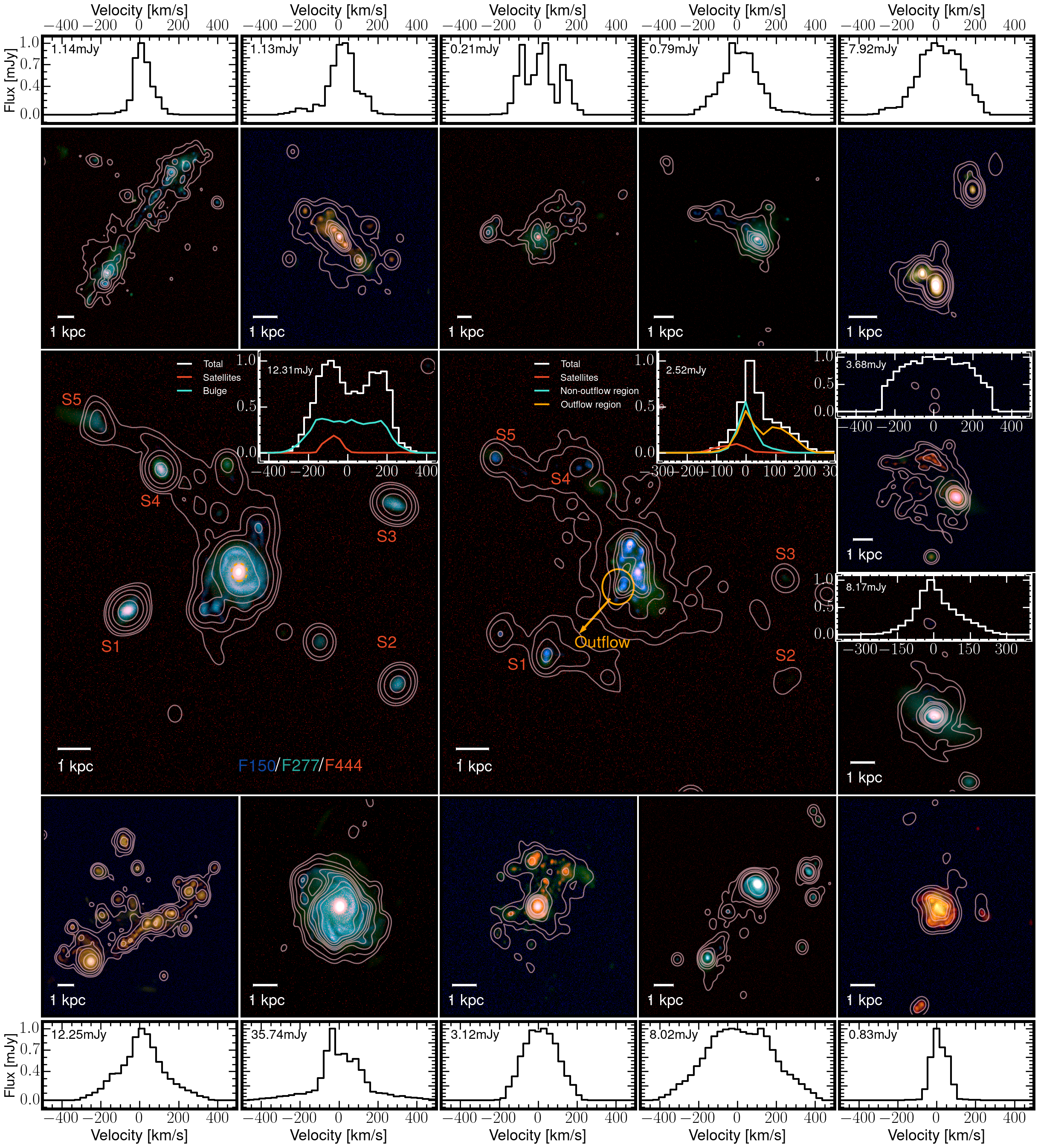}
    \caption{JWST composite images in F150 (b), F277W (g) and F444W (r) overlaid with [C II] line emission contours assuming a 0.3$''$ beam (pink contours) for {\tt SPICE} galaxies at $z=5-7$ for the three feedback models. [C II] contours indicate [2,5,10,20,50,100] $\sigma$ contours assuming a noise of 0.1 mJy. Extended [C II] emission is seen ubiquitously in all models with [C II] extending far beyond the UV component of the galaxies. Insets show normalized [C II] spectra extracted within the effective radius of [C II] for each galaxy, and the numbers listed highlight the peak flux of each spectrum. We see a strong diversity in morphologies (of UV and [C II]) and spectra of [C II] emitters. The middle enlarged panels show the exact same halo in {\tt smooth-sn} (left) and {\tt bursty-sn} (right) at $z=5$, highlighting two distinct mechanisms to produce broad wings in [C II] spectra of galaxies. Galaxy in {\tt smooth-sn} produces a symmetric broad wing due to the presence of a rotating bulge, while in {\tt bursty-sn} this is due to the presence of an outflowing region. This highlights the diverse nature of kinematics traced by [C II] in the first galaxies. 
    }
    \label{fig:UVCIIcontours}
\end{figure*}
As the [C II] emission is excited predominantly by collisions of \textsc{C II} atoms with electrons and \textsc{HI} \citep{Goldsmith2012},
the luminosity emitted by a gas cell with volume $V$ is written as 
\begin{equation}
    L_{\rm CII} =  \sum^{e^-,\rm HI}_i n_i n_{\rm CII}, \Lambda_i (T) V,
    \label{eq:LCII}
\end{equation}
where $\Lambda_i$ is the cooling rate as a function of gas temperature for species $i$ (see Appendix A, also \citealt{Ferrara2019CII} for details). 
Additionally, while [C II] emission remains optically thin for most galaxies and ISM phases \citep{1992Osterbrock}, optical depth within the centers of galaxies can reach moderate levels and lead to self-absorption of [C II] \citep[e.g.][]{2020Guevara}. We model the optical depth in each gas cell as
\begin{equation}
    \tau_{\nu_0} = \frac{c^3}{8 \pi \nu_0^3}A_{\rm ul} \frac{g_l}{Z(T_{\rm ex})} N_{\rm C^+} \biggl[1 - {\rm exp}\biggl(-\frac{h\nu_0}{k T_{\rm ex}} \biggr)\biggr] \frac{1}{\Delta \nu},
\end{equation}
where $\nu_0 = \rm 1900.5 GHz$ is the rest frequency, $A_{\rm ul} = 2.3\times10^{-6}$~s$^{-1}$ is the spontaneous emission coefficient, $g_l=2$ and $g_u=4$ are the level degeneracies, $\mathcal{Z}(T_{\rm ex}) = g_l + g_u {\rm exp}(-91.2/T_{\rm ex})$ is the partition function, $N_{C^+}$ is the column density of $C^+$, $\Delta \nu$ is the doppler width and $T_{\rm ex} = 91.2 / {\rm ln}\bigl[2 \bigl( 1 +3000/n_H \bigr) \bigr]$ is the excitation temperature for the transition.
We find that self-absorption is relevant in galaxies with $M_*>10^9 M_\odot$ which can lead to reduction in luminosity by a factor of $2-5$ \citep[similarly to][]{Katz_2022CII}. 

Finally, at high redshift the cosmic microwave background (CMB) represents a background against which [C II] emission is detected \citep{da_Cunha_2013,Vallini_2015}. If the contrast of [C II] emission against the CMB is not high enough, the [C II] can be obscured. This predominantly affects low density regions \citep{Kohandel2019Kinematics}. We estimate ``obscuration'' factor and scale luminosities in each gas cell (eq~\ref{eq:LCII}) as detailed in \cite{Vallini_2015}.

\begin{figure*}
    \centering
    \includegraphics[width=\linewidth]{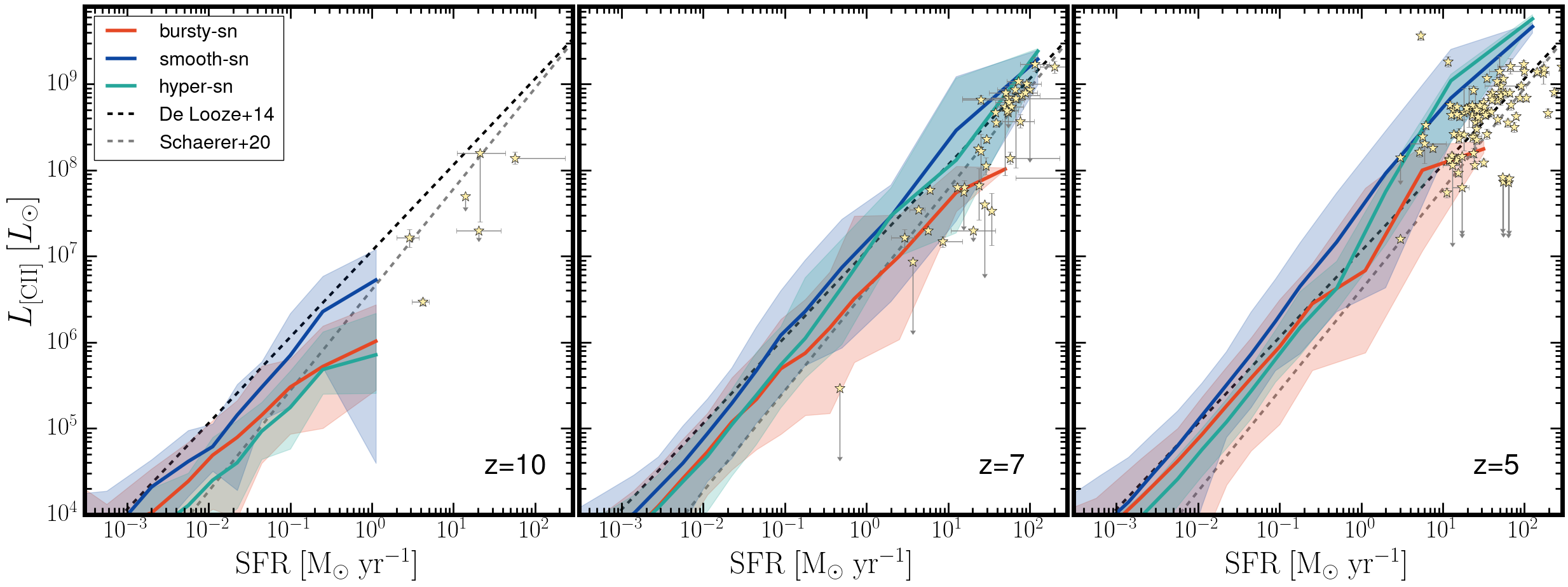}
    \caption{Luminosity of the [C II] line ($L_{\rm [CII]}$) as a function of the SFR calculated over 10 Myr of lookback time for {\tt SPICE} galaxies at $z=10$ (left panel), $z=7$ (middle panel) and $z=5$ (right panel) for {\tt bursty-sn} (red), {\tt smooth-sn} (blue) and {\tt hyper-sn} (turquoise), respectively. While {\tt smooth-sn} and {\tt hyper-sn} exhibit a clear linear trend, {\tt bursty-sn} shows a relative suppression in luminosity at $\rm SFR > 1 M_\odot yr^{-1}$. Yellow stars show ALMA observations of [C II] emitters in redshift bins of 0.5 around the redshift of each panel (see text). Black and grey lines represent best fits from observations of [C II] emitters at $z\sim$0 \citep{DeLooze2014FIRSFR} and $z\sim5-7$ \citep{Schaerer_2020}. Shaded regions enclose 16th and 84th percentiles. Yellow stars show data from various ALMA surveys and serendipitous observations \citep{Capak2015,2013Finkelstein,Ota2014CIIdeficit,Schaerer2015,Maiolino2015CIIdeficit,Pentericci2016CIIdeficit,Knudsen2016CIIdeficit,Laporte2017CIIdeficit,Bradač_2017CIIdeficit,Smit2018,2019Matthee,Bakx2020,Fudamoto_2022,2022Molyneux,Schouws2023,Heintz2023,fudamoto2023ext}. 
    }
    \label{fig:LCIISFR}
\end{figure*}
\subsection{Synthetic observations}
\label{sec:SyntheticdataCII}
\subsubsection{Constructing ALMA datacubes}
We generate synthetic 3D datacubes for each {\tt SPICE} halo to emulate ALMA observations and extract [C II] properties. Each datacube is centered on the center of mass of a halo and spans a fixed field of view of 50 kpc per side. The spatial dimensions are sampled on a grid of $N_{\rm pix} \times N_{\rm pix}$ pixels, with a spatial resolution of $\Delta x \approx 50$ pc, while the spectral (velocity) axis extends from $-1000$ to $+1000$ $\rm km/s$, divided into $N_{\rm chan}$ channels of width $\delta v = 30$ $\rm km/s$. The systemic velocity, $v_{\rm sys}$, is defined as the channel corresponding to the peak of the integrated emission.

Each gas cell contributes a Gaussian spectral profile to the datacube, based on its position and line-of-sight velocity, computed as $v_{\rm los} = \vec{v}_i \cdot \vec{r}$, where $\vec{v}_i$ is the Cartesian velocity of the gas particle and $\vec{r}$ is the observer’s line-of-sight vector.  The luminosity of each particle can be converted to flux density as follows
\begin{equation*}
    I_{i}(v) = \frac{L_{v,i}}{4 \pi d^2 \Delta \nu},
\end{equation*}
where $L_{\nu,i}$ is the spectral luminosity in $\rm erg/s$, $d$ is the luminosity distance, and $\Delta \nu$ is the bandwidth corresponding to a single velocity channel. The line profile contributed by each gas cell is modeled as:
\begin{equation}
I(v) = \frac{I_i}{\sqrt{2\pi}\sigma_v} \exp\left( -\frac{(v - v_{\text{los}, i})^2}{2\sigma_v^2} \right),
\end{equation}
where $\sigma_v$ is the line width, incorporating both thermal broadening, $\sigma_{\rm th} = \sqrt{2kT/m}$, and turbulent broadening, $\sigma_{\rm turb}$, as recorded in the {\tt SPICE} snapshots.

To simulate ALMA’s spatial resolution, we convolve each velocity channel with a 2D symmetric Gaussian beam characterized by a full width at half maximum (FWHM) of $0.3\arcsec$ \citep{Herrera_Camus_2021}. The beam’s standard deviation is given by $\sigma_{\rm beam} = {\rm FWHM}/\sqrt{8\ln 2}$. Additionally, Gaussian random noise is added to each voxel, with an RMS value of $\sigma_{\rm noise} = 0.1$ $\rm mJy/beam$, consistent with typical ALMA sensitivities at $z > 4$ \citep{2020Ginolfi,Fujimoto_2019, 2025Herrera-Camus}.

\subsubsection{Extracting [C II] properties from datacubes}
\label{subsec:CIIdata}
We extract moment -0,1 and 2 maps (which represent integrated intensity, velocity and velocity dispersion respectively) from the datacubes to quantify the morpho-kinematic properties of {\tt SPICE} halos. The moment maps are calculated as: 
\begin{equation}
    M_0 = \sum I(v_i), 
\end{equation}
\begin{equation}
    M_1 = \frac{\sum v_i I(v_i)}{\sum I(v_i)}, \hspace{0.2in} {\rm and}
\end{equation}
\begin{equation}
    M_2 = \sqrt{ \frac{ \sum I(v_i) \cdot (v_i - M_1)^2}{\sum I(v_i)}}.
\end{equation}

We fit 2D Gaussians to the moment-0 maps and use the fitted FWHM to estimate the effective radius of [C II] emission as $R_{\rm [C II]} = 1.21 \times {\rm FWHM} \sqrt{q}$, where ${\rm q}$ is the ratio of major and minor axes of the fitted gaussian. All physical quantities as shown in the following sections are estimated within $R_{\rm [C II]}$, unless otherwise specified. We extract the [C II] line spectrum from a circular aperture with a radius of $R_{\rm [CII]}$, as described previously, and fit Gaussian profiles to the spectral line. The general form of the Gaussian fitting function used is
\begin{equation}
    G(v) = A \cdot e^{-\frac{(v - \mu)^2}{2 \sigma^2}},
\end{equation}

where $A$ is the amplitude, $\mu$ is the line center, and $\sigma$ is the line width. Each spectrum is fit with up to three Gaussian components, and the optimal number of components is determined using the Bayesian Information Criterion (BIC), defined as
\begin{equation}
    \mathrm{BIC} = \chi^2 + k\ln(n),
\end{equation}
where $\chi^2$ is the chi-squared value for the model, $k$ is the number of free parameters and $n$ is the number of datapoints used in the fit. The model yielding the lowest BIC is selected as the best fit. For all Gaussian components, the following parameter bounds are imposed: $\mu \in [-300, 300]$ $\rm km/s$, $\sigma \in [0, 1000]$ $\rm km/s$, and $A \in [0, \max(I_\nu)]$ $\rm mJy$, where $\max(I_\nu)$ denotes the peak flux of the extracted spectrum. When two Gaussian components are fitted, we explore the efficacy of classically used techniques to estimate outflow velocities \citep[e.g.][]{Herrera_Camus_2021}, where the outflow velocity is defined as $v_{\rm out} \sim |\Delta v_{\rm broad}| + \rm FWTM/2$ \citep[e.g.][]{Lutz_2020}, with $\Delta v_{\rm broad}$ shift of the line center of the broad Gaussian component with respect to the systemic velocity, and $\rm FWTM$ full-width at a tenth of the maximum of the broad component.

Kinematic properties such as the rotational velocity ($V_{\rm rot}$) and velocity dispersion ($\sigma_0$) are modeled via two methods: (i) intrinsic 3D rotation and dispersion curve using {\tt SPICE} snapshots; (ii) calculating $V_{\rm obs}$ and $\sigma_{\rm int}(\rm [CII])$ (derived from single Gaussian fit to the spectrum) as described in \citet{Lilian2025}, who model observations of [C II] emitters at $z>4$. We use these methods to estimate $V/\sigma$ values to understand the impact of stellar feedback on the formation of dynamically cold disk galaxies in {\tt SPICE}. We classify galaxies with $V_{\rm rot}/\sigma_0 > \sqrt{3.36}$~ ($V_{\rm obs}/2\sigma_{\rm int} >$ 0.4) as disk galaxies \citep{Schreiber_2018}, while the remainder are classified as dispersion dominated galaxies. Finally, we estimate the effective size of UV\footnote{UV emission here refers to integrated emission between 1400-2600 \AA.} emission \citep[see][]{BhagwatLya_2024} assuming the observational depth of the JADES survey \citep{eisenstein2025}.
\section{Results}
\label{sec:resultsCII}

In this section, we present our main results on the properties of [C II] emission in the {\tt SPICE} galaxy sample at $z=5$–10. We focus in particular on the spatial distribution and morphology of the [C II] emitting gas, its connection to star-forming regions traced in the rest-frame UV, and the kinematic signatures encoded in the [C II] line profiles. These diagnostics allow us to explore how [C II] traces the structure of star-forming gas and dynamical processes such as rotation, turbulence, and outflows operating in the first galaxies.
\subsection{Overview}
\label{subsec:overview}

Figure~\ref{fig:UVCIIcontours} provides an overview of these properties by comparing mock JWST imaging with simulated [C II] emission for {\tt SPICE} galaxies across three feedback models. The JWST composite images combine F150W (blue), F277W (green), and F444W (red) bands, while pink contours indicate [C II] line emission assuming a $0.3''$ beam. The contours correspond to $[2,5,10,20,50,100]\sigma$ levels, adopting a noise level of 0.1 mJy. In all feedback models, [C II] emission is spatially extended and frequently reaches well beyond the UV-bright regions of the galaxies, indicating that a substantial fraction of the [C II] emitting gas resides outside the compact stellar component.

The diversity of [C II] morphologies and spectral profiles across the sample is illustrated by the inset panels, which show normalized [C II] spectra extracted within the effective radius of the [C II] emission for each galaxy. Both the spatial distribution of [C II] and the spectral shapes vary significantly among systems, reflecting a wide range of gas distributions and dynamical states in the simulated galaxies.

The enlarged middle panels highlight this diversity by showing the same halo at $z=5$ in two feedback models: {\tt smooth-sn} and {\tt bursty-sn}. The [C II] spectra differ markedly due to the distinct feedback-driven gas dynamics. In the {\tt smooth-sn} case, the galaxy exhibits a symmetric broad component produced by the presence of a rotating bulge. In contrast, the {\tt bursty-sn} model shows broad spectral wings associated with an outflowing gas component. These examples illustrate that broad [C II] line features can arise from multiple physical mechanisms, underscoring the complex and diverse kinematics traced by [C II] in the first galaxies

\subsection{Luminosity of the [C II] line}
\label{subsec:luminosity}

\begin{figure}
    \centering
    \includegraphics[width=0.928\linewidth]{Plots/fHIvsLCII.png}
    \caption{Median volume filling fraction of HI within the effective radius of [C II] emission (see section~\ref{subsec:radcii}) as a function of the [C II] luminosity for {\tt SPICE} galaxies at $z=10$ (top panel), $z=7$ (middle panel) and $z=5$ (bottom panel) for {\tt bursty-sn} (red), {\tt smooth-sn} (blue) and {\tt hyper-sn} (turquoise). Shaded regions enclose 16th and 84th percentiles.}
    \label{fig:fHIvsLCII}
\end{figure}

The [C II] line is considered to be an excellent tracer of star-formation, and observations in the local universe show a very tight correlation between $L_{\rm [CII]}$ and SFR of a galaxy \citep{2025Herrera-Camus,DeLooze2014FIRSFR,Schaerer_2020}. However, at higher redshifts, the scatter around the expected correlation increases \citep{2018LagacheL,Casavecchia_2024,Lilian2025}. The origin of this increased scatter remains debated, and previous studies hint toward feedback playing a key role. In this context, we first explore the connection between the luminosity of the [C II] line and SFR of {\tt SPICE} galaxies.

In Figure~\ref{fig:LCIISFR}, we show $L_{\rm [CII]}$ as a function of the SFR of a galaxy at $z=10-5$ for the three feedback models. Luminosities predicted for the {\tt SPICE} galaxies are in good agreement with ALMA observations of emitters from large surveys and serendipitous detections (see caption). Already at $z=10$ we see that a "main-sequence" is established in the $L_{\rm CII}-\rm SFR$ plane for all three models, which shows a minimal evolution down to $z=5$. At $z=10$, the {\tt smooth-sn} model predicts the brightest [C II] luminosities, exceeding the other models by 0.2–1 dex, while {\tt bursty-sn} and {\tt hyper-sn} remain comparable. At $z=7$, the three models are indistinguishable at $\rm SFR < 10^{-1} \ M_\odot yr^{-1}$. At $\rm SFR > 10^{-1}\, M_\odot\,yr^{-1}$, {\tt smooth-sn} and {\tt hyper-sn} predict similar [C II] luminosities, with relatively small scatter. In contrast, {\tt bursty-sn} exhibits a pronounced deficit\footnote{Not to be confused with deficit of $L_{\rm [CII]}$ in relation to infrared luminosity $L_{\rm FIR}$.} of $0.7$--$1$ dex as compared to the other models, and a much larger scatter, deviating from the tight $L_{\rm [CII]}$--SFR relation. A similar trend is seen at $z=5$.

The relative suppression of $L_{\rm [CII]}$ in {\tt bursty-sn} is particularly interesting as it has been observed also in other studies that employ different numerical models and simulation codes. \citet{Katz_2022CII} find that in the {\tt SPHINX} simulations ``bursty leakers'' exhibit $L_{\rm [CII]}$ lower than ``non-bursty leakers'' (see their Figure 11). Similarly, \citet{Casavecchia_2024} and \citet{Liu2026} find a suppression in $L_{\rm [CII]}$ for comparable SFRs using the {\tt ColdSIM}~\citep{Maio_2022} and {\tt FIRE} simulations \citep{Wetzel_2023} respectively. This suggests that the observed luminosity deficit might be connected to the burstiness of feedback, and understanding its origin is key.

\begin{figure}
    \centering
    \includegraphics[width=0.99\linewidth]{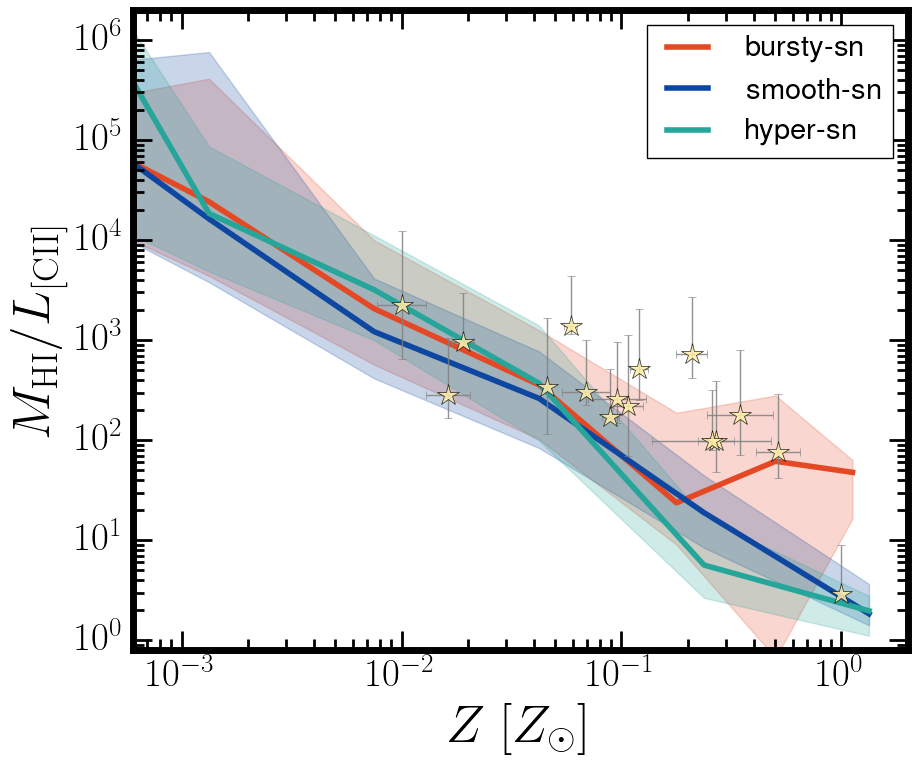}
    \caption{Median mass-to-light ratio $M_{\rm HI}/L_{\rm [C II]}$ as a function of the metallicity of {\tt SPICE} galaxies at $z=5-10$ for {\tt bursty-sn} (red), {\tt smooth-sn} (blue) and {\tt hyper-sn} (turquoise). Shaded regions enclose 16th and 84th percentiles. Yellow data points represent measurements form  \citet{Heintz_2021} and \citet{sulzenauer2025}.}
    \label{fig:MHIvsZ}
\end{figure}

\begin{figure}
    \centering
    \includegraphics[width=0.99\linewidth]{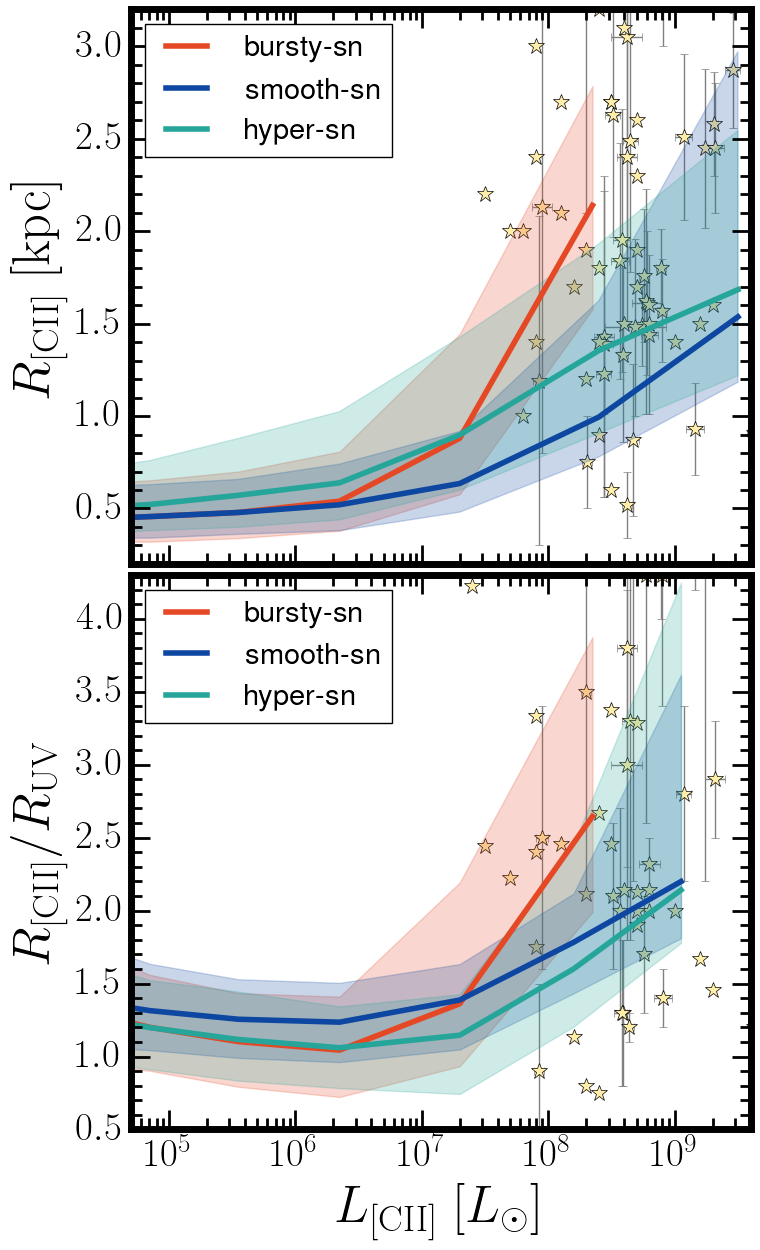}
    \caption{{\it Top:} Median [C II] effective radius as a function of the [C II] luminosity for {\tt SPICE} galaxies at $z=10-5$ for {\tt bursty-sn} (red), {\tt smooth-sn} (blue) and {\tt hyper-sn} (turquoise).
    {\it Bottom:} Median ratio of effective radius of [C II] to UV continuum at 1500~\AA~as a function of [C II] luminosity ($L_{\rm [C II]}$) for {\tt SPICE} galaxies at $z=10-5$ for {\tt bursty-sn} (red), {\tt smooth-sn} (blue) and {\tt hyper-sn} (turquoise).
    Shaded regions enclose 16th and 84th percentiles. Yellow stars show size estimates taken from various large ALMA programs \citep{Fujimoto_2020,Fudamoto_2022, 2024IkedaI, rowland2024rebels25, Bouwens2022Rebels}.}
    \label{fig:RCIIRUVvsLCII}
\end{figure}

As [C II] emission is theorized to originate mainly from PDRs within the ISM, an estimate of the fraction of gas in a galaxy which is able to emit [C II] could be obtained by evaluating the fraction of gas hosting PDRs. Their abundance is shown to be sensitive to feedback, as this alters the porosity and ionisation state of the ISM \citep{Cormier_2019,Madden_2020,gurman2024ci}, and thus we expect to obtain different PDR distributions and abundances in our models. Since in cosmological simulations resolving individual PDRs is not feasible due to the extreme resolution demands, in this work we assume that a cell hosts a PDR, depending on the local ionisation state and H~I fraction of gas \citep{Olsen_2017,Katz_2022CII} (typically $T<10^4$ K, $x_{\rm HI}$ $\approx 1$). 

In Figure~\ref{fig:fHIvsLCII} we show the volume filling fraction of HI ($f_{\rm HI} = \Sigma V_i x_{HI} / \Sigma V_i$) calculated with $R_{\rm eff}$ (see section~\ref{subsec:radcii}) as a function of $L_{\rm [CII]}$ for galaxies at $z=5-10$. At all redshifts, the three models exhibit a $f_{\rm HI}$ decreasing with increasing [C II] luminosity. At $z=10$ all models show similar H~I filling fractions, with {\tt smooth-sn} having the highest $f_{\rm HI}$ at $L_{\rm [CII]} > 10^5$ L$_\odot$. Similarly, at $z=7$, {\tt smooth-sn} exhibits the highest $f_{\rm HI}$ followed by {\tt hyper-sn}. Meanwhile, {\tt bursty-sn} shows a strong suppression (0.2-1 dex) in $f_{\rm HI}$ above $L_{\rm [CII]} > 10^6$ L$_\odot$. A similar trend can be seen also at $z=5$.  The strong drop in $f_{\rm HI}$ for {\tt bursty-sn} indicates a higher level of ionisation and a lower gas surface density, which is a direct consequence of feedback  \citep[see][]{bhagwat2024}. This is consistent with previous findings \citep{asada2023bursty,Katz_2022CII,Casavecchia_2024}. The reduction in $f_{\rm HI}$ leads to a lower total fraction of PDRs which emit [C II], and thus to the differences observed in the $L_{\rm CII} - \rm SFR$ relation.

In addition to tracing the SFR of a galaxy, the [C II] luminosity can also be used to estimate its total atomic gas budget \citep{Herrera_Camus_2021,2025Casavecchia}, and various studies aim to use a mass-to-light ratio to estimate the galactic H~I mass from the [C II] luminosity \citep{Casavecchia2025}.
In Figure~\ref{fig:MHIvsZ}, we present the mass-to-light ratio as a function of [C II] luminosity–weighted metallicity for {\tt SPICE} galaxies at $z=5$–10. The scaling against metallicity is key as [C II] emissivity depends on the gas phase metallicity (see Eq. \ref{eq:LCII}). The three models are indistinguishable at $Z < 0.2$ Z$_\odot$. At higher metallicities ($Z > 0.2$ Z$_\odot$), however, the {\tt bursty-sn} model exhibits a flattening and systematically higher values compared to the other models. This implies that at these metallicities, galaxies with bursty feedback require more H~I to produce the same [C II] luminosity as galaxies with a smooth feedback, and thus, at fixed [C II] luminosity, the conversion factor to $M_{\rm HI}$ is larger in the former scenario than in the latter. Combined with Figure~\ref{fig:fHIvsLCII}, which shows that bursty galaxies are H~I depleted, we are able to explain the deficit in [C II] luminosity for these galaxies. 
When comparing to observational data, the {\tt bursty-sn} model provides the best match to estimates using absorption line measurements against gamma-ray bursts at $z=2-6$ \citep{Heintz_2021}, in which $M_{\rm HI}$ is evaluated from H~I absorption while $L_{\rm [C II]}$ from [C I$\rm I^*$]$\lambda$ 1335 absorption. In contrast, the {\tt smooth-sn} and {\tt hyper-sn} models predict significantly lower mass-to-light ratios, consistent with recent observations of 10 galaxies in the SPT2349-56 protocluster at $z=4.3$ \citep{sulzenauer2025}, which report an exceptionally low mass-to-light ratio of $2.9 \pm 0.3$.

From the discussion above we can conclude that feedback in {\tt SPICE} regulates the cold gas content in a way that suppresses the [C II] emission in bursty star-forming galaxies. As observations from large ALMA surveys such as ALPINE have presented non-detections of [C II] in massive galaxies at $z>5$ \citep{Romano2022CIIdeficit,Romano_2022}, we argue that such non-detections could potentially act as a tracer for bursty feedback.

\subsection{Spatial distribution of [C II] emission}
\label{subsec:radcii}
Previously, we quantified the luminosity of [C II] and its connection to the HI mass budget of galaxies. Here, we characterize the spatial distribution of [C II] emitting gas. In the top panel of Figure~\ref{fig:RCIIRUVvsLCII} we show the size-luminosity relation for {\tt SPICE} galaxies at $z=5-10$\footnote{We do not find a notable redshift evolution in observed properties unless otherwise mentioned.}, where the size is characterized in terms of the effective radius $R_{\rm [CII]}$ defined in section~\ref{subsec:CIIdata}. In galaxies with $L_{\rm [CII]} \lesssim 10^{6.5}$ L$_\odot$, the three models converge, predicting compact emission such that the radius exhibits only a marginal increase with luminosity. For $L_{\rm [CII]} > 10^7$ L$_\odot$, all models show a pronounced increase in effective radii, with the most luminous systems also being the most extended. Among the models, {\tt bursty-sn} typically yields the largest radii due to efficient feedback driving gas outward which also leads to lowering of central concentration of gas, followed by {\tt hyper-sn}, while {\tt smooth-sn} produces the most compact [C II]-emitting galaxies due to largely inefficient feedback. Datapoints overplotted show measurements from various [C II] observations at $z=4-8$ \citep{Carniani_2018,Fudamoto_2022,2024IkedaI}. While the predicted [C II] sizes agree well with observations, one cannot distinguish between models due to the scatter in the data.

\begin{figure}
    \centering
    \includegraphics[width=0.99\linewidth]{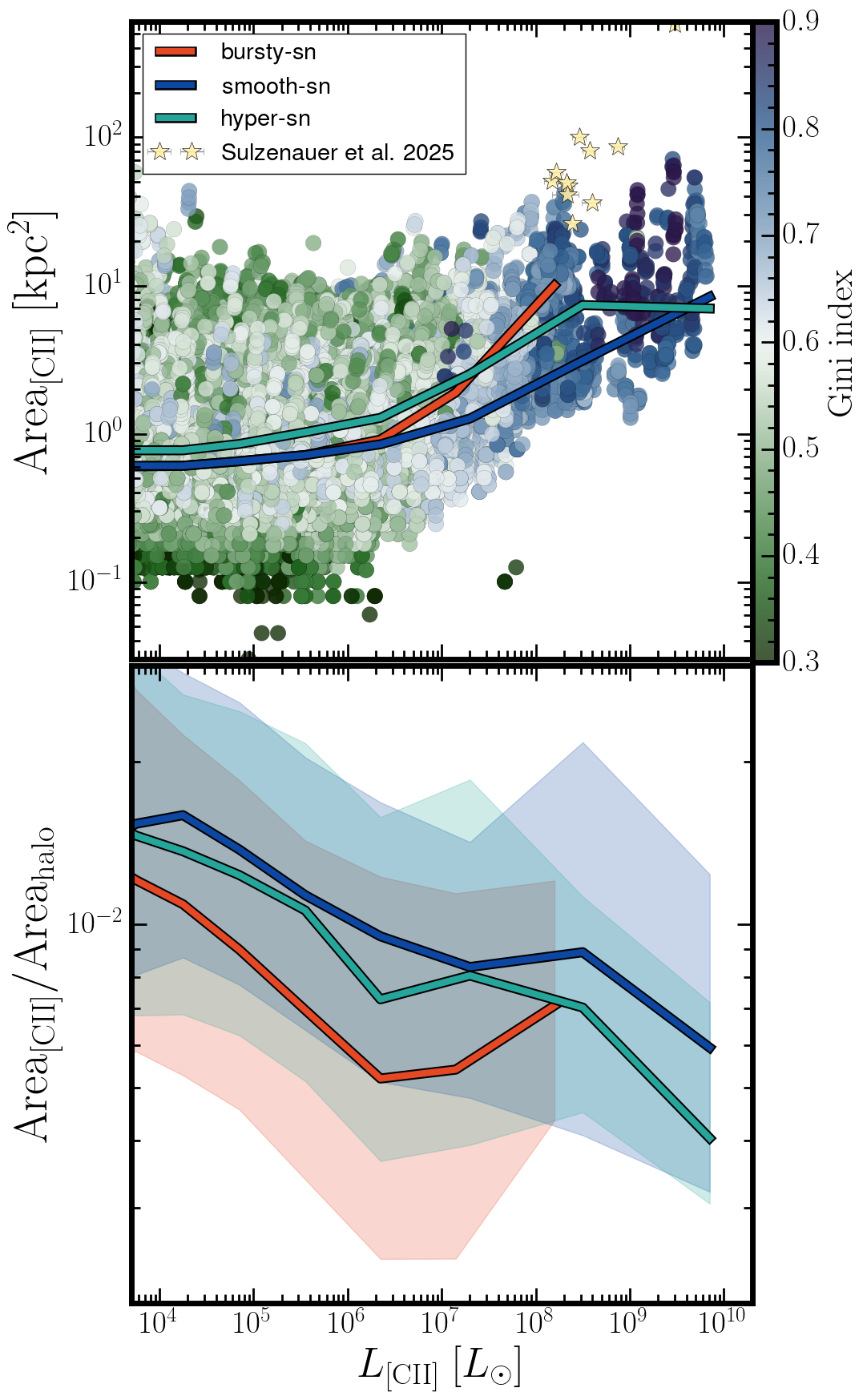}
    \caption{{\textit{Top:} Area of [CII] emission enclosed within the $2\sigma$ contour as a function of the [C II] luminosity for {\tt bursty-sn} (red), {\tt smooth-sn} (blue) and {\tt hyper-sn} (turquoise) at $z=10-5$. Solid lines show medians for the respective models while the scatter includes galaxies from all three models. Each scatter point is also color coded with the Gini index estimated for the emission. Datapoints show size estimates for [C II] emitting regions in the SPT2349-56 protocluster from \citealt{sulzenauer2025}.}
    {\textit{Bottom:}} Ratio of area of [C II] emission to the area of the halo as a function of [C II] luminosity for {\tt bursty-sn} (red), {\tt smooth-sn} (blue) and {\tt hyper-sn} (turquoise). Shaded regions enclose 16th and 84th percentiles.}
    \label{fig:AreaLum}
\end{figure}

The ratio of the effective radius of [C II] to the rest-frame UV continuum is often used to ascertain the presence of an extended [C II] ``halo'' which indicates extent and strength of metal enrichment in the CGM. In the bottom panel of Figure~\ref{fig:RCIIRUVvsLCII} we show such ratio (see section~\ref{sec:SyntheticdataCII}) as a function of the [C II] luminosity of {\tt SPICE} galaxies at $z=5-10$. At all luminosities, the [C II] emission is systematically more extended than the UV continuum ($R_{\rm [CII]}/R_{\rm UV}$ > 1). The relative spatial extent of [C II] shows a lack of evolution up to $L_{\rm [CII]} \approx 10^{6.5} L_\odot$, above which the ratio increases progressively with [C II] luminosity. At a fixed luminosity for $L_{\rm [CII]} \gtrsim 10^{6.5}$ L$_\odot$, {\tt bursty-sn} exhibits the largest $R_{\rm [CII]}/R_{\rm UV}$, with a median of $2.2\pm0.2$, followed by {\tt smooth-sn} and {\tt hyper-sn} with $1.8\pm0.2$ and $1.7\pm0.4$, respectively. Although the predictions from {\tt SPICE} are consistent with current observations \citep{Carniani_2018,Fudamoto_2022,2024IkedaI}, the ratio $R_{\rm [CII]}/R_{\rm UV}$ provides limited discriminatory power between the feedback models due to the large scatter in the data. 

While the effective radius of [C II] represents a circularized radius, in reality the emission can be asymmetric and dependent on the spatial distribution of gas and its ionisation state (e.g~Fig~\ref{fig:UVCIIcontours}). The projected area of emission (calculated as ${\rm Area}_{\rm [C II]} = \pi R_{\rm [C II]}^2$) provides a measure of the total surface over which [C II] emission occurs. In the top panel of Figure~\ref{fig:AreaLum}, we show ${\rm Area}_{\rm [C II]}$ as a function of [CII] luminosity. The area of [C II] emission remains constant with increasing luminosity up to $L_{\rm [C II]} \approx 10^{6.5}$ L$_\odot$, above which we see a sharp increase in area with increasing luminosity for {\tt bursty-sn}, and a gradual increase for {\tt smooth-sn} and {\tt hyper-sn}. Additionally, in the top panel of the Figure we show data points from all feedback models combined, color coded by the Gini index which is a measure of clumpiness \citep{1912Gini}\footnote{The Gini index quantifies the inequality of a distribution, ranging from 0 (uniform) to 1 (maximally unequal).}, estimated for the [C II] emission using the moment-0 maps. We see that the brightest [C II] emitting galaxies tend to exhibit the highest Gini indices, with the Gini index increasing progressively with increasing luminosity irrespective of the feedback model. This indicates that [C II] luminosity in the brightest emitters ($L_{\rm [C II]} \geq 10^{8}$ L$_\odot$) originates from a number of small compact emitting regions, while for lower luminosities the luminosity emerges from a more uniform distribution of gas. 
\begin{figure}
    \includegraphics[width=0.99\linewidth]{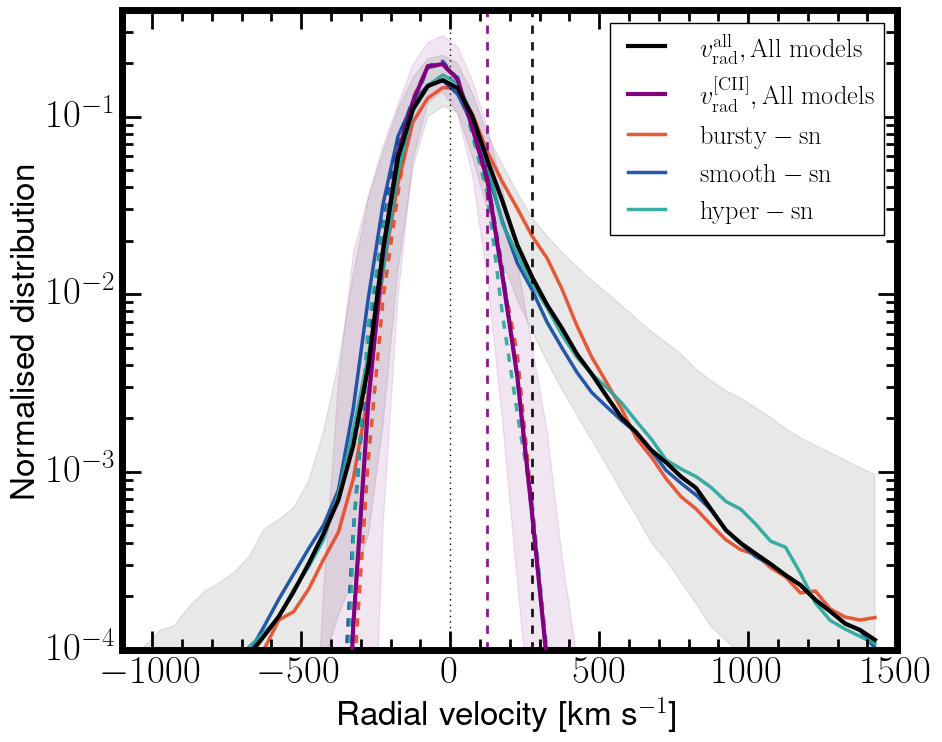}
    \caption{Median radial velocity distributions of gas in halos hosting galaxies with $M_* > 10^8\,\mathrm{M_\odot}$. The black curve shows the median distribution computed over all gas cells, combining all feedback models, with the shaded region indicating the $16$--$84$ percentile range. The purple curve corresponds to [C II]-bright gas ($L_{\rm [CII]} > 0.01\,\mathrm{L_\odot}$). Colored solid (dashed) curves show the distributions for all gas cells ([C II] bright) for individual feedback models. $v_{\rm rad} > 0$ denotes outflowing gas and $v_{\rm rad} < 0$ denotes inflowing gas. Vertical dashed purple and black lines represent the $\rm 90^{th}$ percentiles of the distributions.}
    \label{fig:vradall}
\end{figure}
\begin{figure}
    \includegraphics[width=0.99\linewidth]{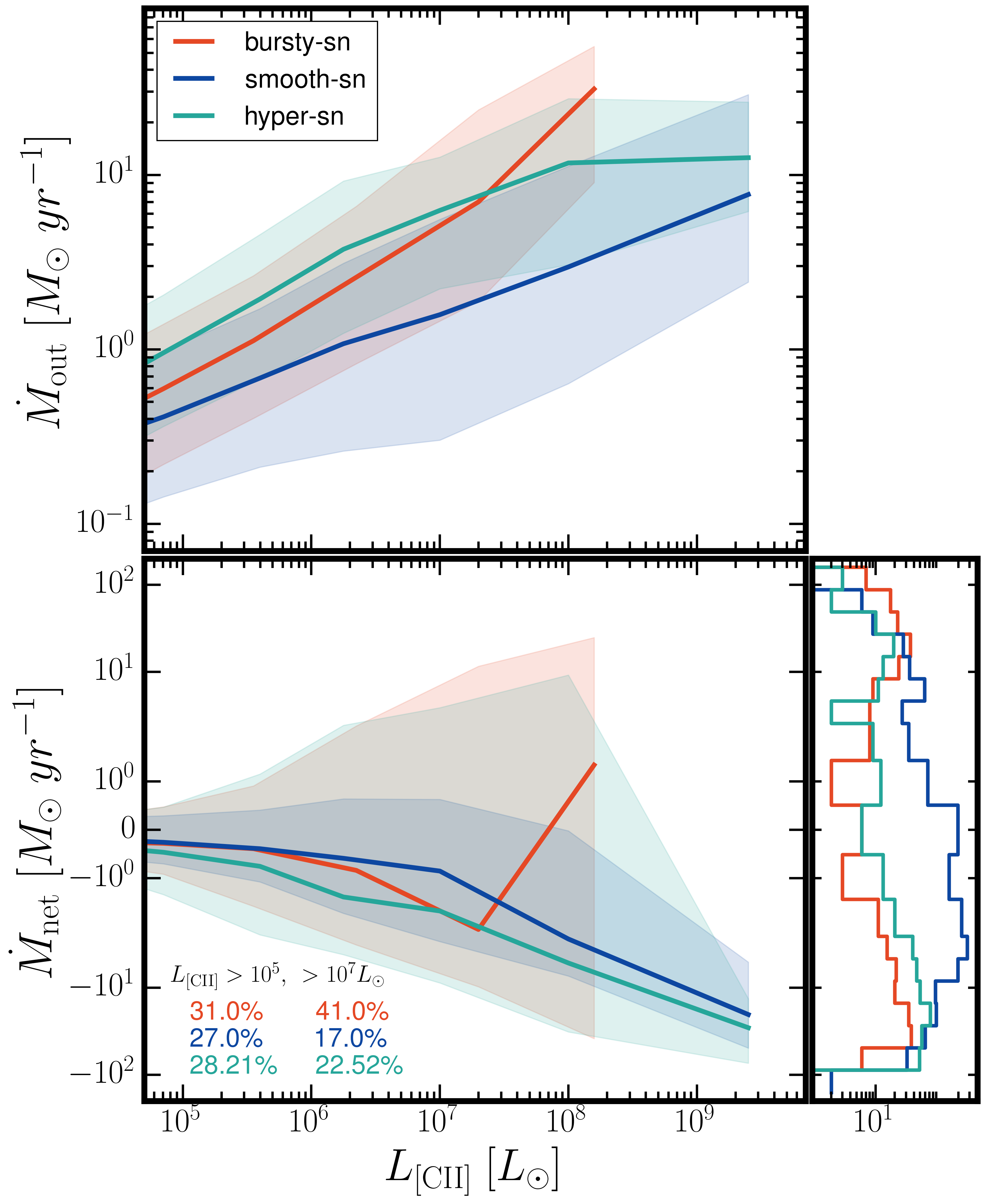}
    \caption{ {\it Top:} Median mass outflow rates for the three feedback models as a function of the [C II] luminosity, $L_{\mathrm{[CII]}}$. Shaded regions denote the 16th and 84th percentiles. {\tt bursty-sn} transitions from net inflow to net outflow at high $L_{\rm [C II]}$, driving the strongest outflows, whereas {\tt smooth-sn} and {\tt hyper-sn} demonstrate lower outflow efficiencies with increasing luminosities. 
    {\it Bottom:} 
    Median net mass flow rate, $\dot{M}_{\mathrm{net}}$, as a function of [C II] luminosity, $L_{\mathrm{[CII]}}$, for the \texttt{bursty-sn} (red), \texttt{smooth-sn} (blue),  and \texttt{hyper-sn} (turquoise) feedback models. Shaded regions denote the 16th and 84th percentiles. The histograms display the distributions of $\dot{M}_{\mathrm{net}}$ for each model for galaxies with $L_{\rm [C II]} > 10^7 L_\odot$. Numbers listed indicate fraction of galaxies that are outflow dominated at different luminosity cuts. 
    }
    \label{fig:mnetvsLCII}
\end{figure}
\begin{figure}
    \centering
    \includegraphics[width=0.95\linewidth]{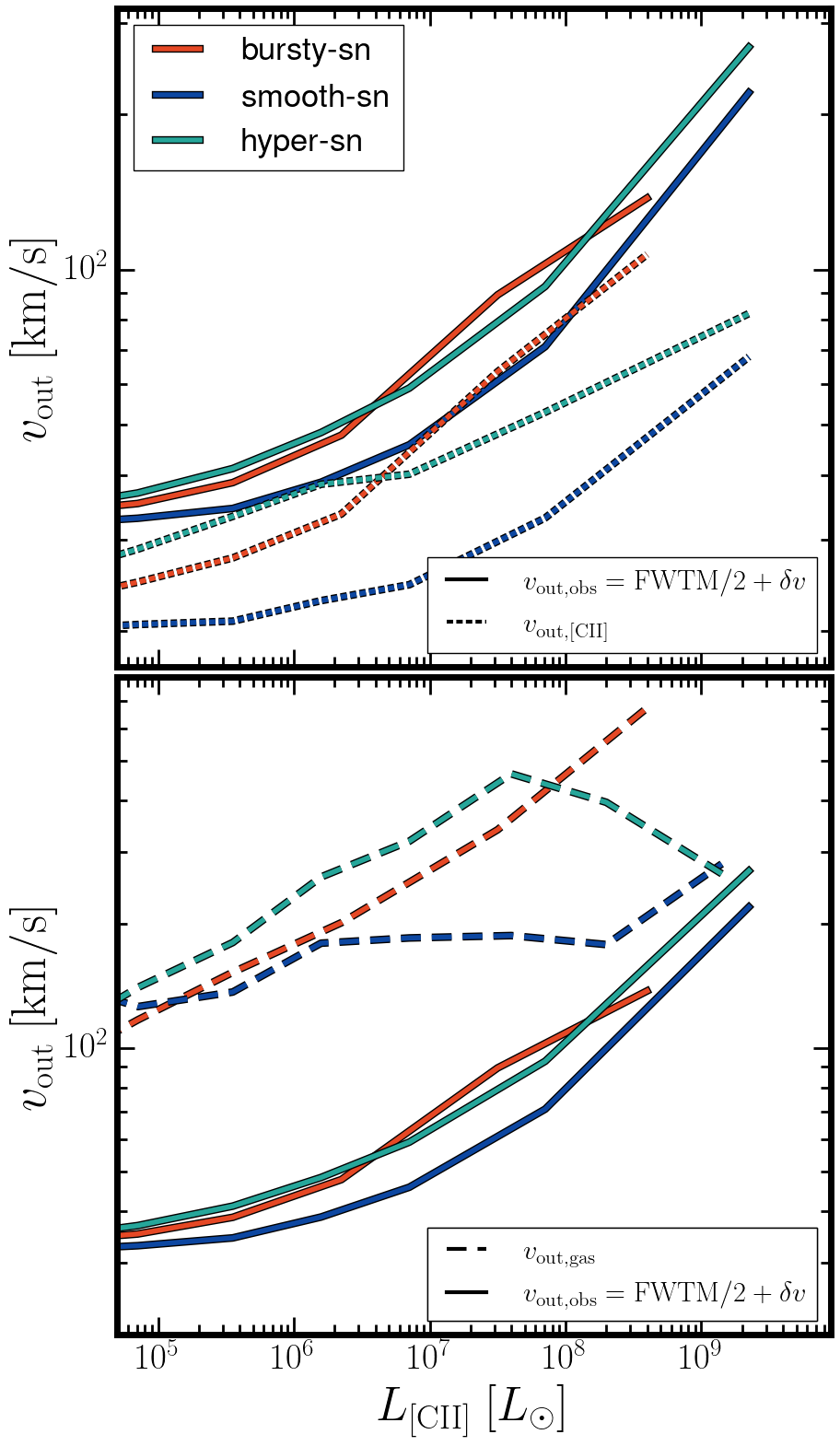}
    \caption{ ({\it Top}): Median outflow velocities of [C II]-bright (cold) gas derived from the simulations ($v_{\rm out, [CII]}$, dotted lines) and those inferred from [C II] spectra ($v_{\rm out, obs}$, solid lines) as a function of $L_{\rm [C II]}$ for {\tt SPICE} galaxies at $z=10$–$5$, shown for the {\tt bursty-sn} (red), {\tt smooth-sn} (blue), and {\tt hyper-sn} (turquoise) feedback models. ({\it Bottom}): Median outflow velocities of all gas phases combined ($v_{\rm out, gas}$, dashed lines) and those inferred from [C II] spectra ($v_{\rm out, obs}$, solid lines) as a function of $L_{\rm [CII]]}$. Overall, [C II]-based kinematic estimates tend to underestimate the true outflow velocities of all gas, while overestimating those of the cold component.}
    \label{fig:voutvssigmaSFR}
\end{figure}

\begin{figure}
    \includegraphics[width=0.96\linewidth]{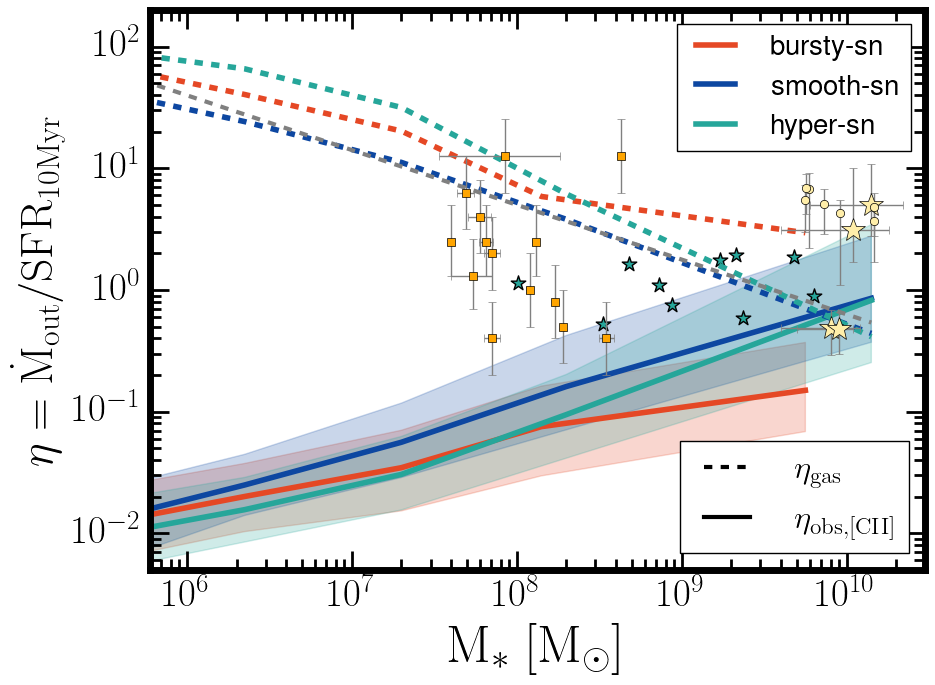}
    \caption{Median mass loading rate calculated at $R_{\rm [CII]}$ as a function of stellar mass of {\tt SPICE} galaxies at $z=10-5$ for {\tt bursty-sn} (red), {\tt smooth-sn} (blue) and {\tt hyper-sn} (turquoise). Dashed colored lines show the estimates using {\tt SPICE}, while solid lines are obtained following an observational style approach. Yellow stars show ALMA \citep{Herrera_Camus_2021,2020Ginolfi,birkin2025alma} constraints from $z>5$ [C II] emitters, orange squares show ionised gas derived values using {\tt JWST} \citep{Carniani_2024}, blue stars show Herschel \citep{Romano_2022} estimates for $z\approx0$ dwarfs galaxies and yellow circles show predictions from semi-analytical models for {\tt ALPINE} galaxies from \citealt{Pizzati_2022outflow} respectively.}
    \label{fig:etavsMstar}
\end{figure}

In the bottom panel of Figure~\ref{fig:AreaLum}, we show the  ratio of area of [C II] emission to the projected area of the host dark matter halo (${\rm Area}_{\rm [CII]}/{\rm Area}_{\rm halo}$) as a function of the [C II] luminosity of the galaxy. We note a strong anti-correlation between the  ratio ${\rm Area}_{\rm [CII]}/{\rm Area}_{\rm halo}$ and the luminosity of the galaxy, implying that the fractional area covered by [C II] emission decreases with increasing [C II] luminosity. Both {\tt smooth-sn} and {\tt hyper-sn} exhibit a power-law decline across all luminosities, whereas {\tt bursty-sn} shows an initial decrease up to $L_{\rm [CII]} \sim 10^6$ L$_\odot$, followed by an upturn beyond $L_{\rm [CII]} \gtrsim 10^7$ L$_\odot$. We note that there are $\approx$410 galaxies in the bins with $L_{\rm [C II]} > 10^7$ L$_\odot$ for {\tt bursty-sn} therefore the turnover is statistical. The {\tt bursty-sn} model generally exhibits the smallest fractional [C II]-emitting area at all luminosities because, at a given halo mass, {\tt bursty-sn} produces lower [C II] luminosities than the other two models (see Fig.~\ref{fig:LCIISFR}). 

\subsection{[C II] as a tracer of gas flows}
\label{subsec:ciiflows}
While we previously analysed the spatial extent and morphology of the [C II]-emitting gas, we now turn to its dynamics and, in particular, its merit as a tracer of galactic outflows. 
Derived quantities, such as mass outflow rate ($\dot{M}_{\rm out}$), outflow velocity ($v_{\rm out}$), and mass-loading factors, are commonly used to quantify the strength of feedback in [C II] emitting galaxies. 
Observationally, broad wings in [C II] spectra are used as proxy for outflows, with the line width and flux used to constrain outflow speed and mass, respectively. 
Here, we investigate how intrinsic outflow properties respond to our different stellar feedback models, comparing them with the outflow properties that would be inferred from synthetic datacubes.
\subsubsection{Radial velocity distribution within {\tt SPICE} galaxies}
\label{subsec:kinematicstruct}
We start by looking at the typical radial velocity structure of {\tt SPICE} galaxies. We adopt the definition of outflowing motion (similarly for mass flux etc.) as gas with radial velocity $v>0 \, \rm km \, s^{-1}$ and inflow as $v<0 \, \rm km \, s^{-1}$. The solid black curve in Figure~\ref{fig:vradall} shows the median radial velocity distribution for all gas within halos hosting galaxies with $M_* > 10^8\,\mathrm{M_\odot}$, including the three simulations. 
The purple curve shows the corresponding distribution for [C II]-bright gas, with $L_{\rm [CII]} > 0.01\,\mathrm{L_\odot}$\footnote{We have verified that the choice of luminosity threshold does not affect the qualitative trends.}. 
The full gas distribution is asymmetric, exhibiting a pronounced tail toward high positive velocities ($\gtrsim 500\,\mathrm{km\,s^{-1}}$). This tail traces supernova-driven outflows, which can be seen to reach velocities $\gtrsim 1500 \, \rm km \, s^{-1}$. In contrast, the [C II]-bright distribution (purple curve) is significantly narrower, and much more symmetric. It has an approximately Gaussian shape, centered at ($v_{\rm rad} \approx -50\,\mathrm{km\,s^{-1}}$), and is sharply truncated at high speeds. Taken at face value, [C II] emission preferentially traces dynamically colder gas with modest radial velocities ($|v_{\rm rad}| \lesssim 200\,\mathrm{km\,s^{-1}}$), failing to capture the high-velocity phase that dominates the wind energetics \citep[e.g.][]{Hu2019}. 

The only slight offset from systemic ($v_{\rm rad}=0 \, \rm km \, s^{-1}$) is consistent with a combination of net inflow, turbulent motions, and recycled gas in fountain flows, rather than coherent high-velocity outflows. Consequently, velocity estimates based on [C II] may be biased toward a near-systemic gas and do not reflect the full velocity structure of multiphase outflows.

\subsubsection{Mass flow rates}
\label{subsec:mdotout}

In Figure~\ref{fig:mnetvsLCII} we show gas mass flow rates, defined as the mass flux flowing inwards or outwards across a shell with a radius $R_{\rm [C II]}$, as a function of the [C II] luminosity of galaxies in the three simulations. The upper panel presents the total mass outflow rate ($\dot{M}_{\mathrm{out}}$), while the lower panel displays the net mass flow rate ($\dot{M}_{\mathrm{net}} = \dot{M}_{\mathrm{out}} + \dot{M}_{\mathrm{in}}$), where $\dot{M}_{\mathrm{in}}$ is negative to denote inflow.
Mass fluxes are obtained for all gas phases (irrespective of $L_{\rm [C II]}$). We evaluate these quantities at $R_{\rm [C II]}$ to reduce contamination to inferred quantities from the turbulent ISM of the galaxy. 

All three simulations produce a clear positive correlation between $\dot{M}{_\mathrm{out}}$ and $L_{\mathrm{[CII]}}$, well described by a power-law (best-fit parameters are provided in Appendix~\ref{sec:appendixA}). Among the models, \texttt{smooth-sn} consistently produces the least mass-loaded outflows across the entire luminosity range. In contrast, \texttt{bursty-sn} and \texttt{hyper-sn} display comparable outflow mass flow rates at low to intermediate luminosities, diverging above $L_{\mathrm{[CII]}} \gtrsim 2\times10^{7}$~L$_\odot$, where \texttt{bursty-sn} exhibits a pronounced upturn, while \texttt{hyper-sn} mildly flattens. 

As seen in the bottom panel of Figure~\ref{fig:mnetvsLCII}, at low luminosities ($L_{\mathrm{[CII]}} \lesssim 10^{7}$ ~L$_\odot$) all three models show similar behaviour, exhibiting negative net $\dot{M}_{\mathrm{net}}$ values, typically between $0$ and $\sim -10 - 0$~M$_\odot\,\mathrm{yr^{-1}}$, indicating a mild net inflowing mass-flux. As $L_{\mathrm{[CII]}}$ increases, {\tt smooth-sn} and {\tt hyper-sn} show a stronger net inflow, reaching values of 10-50~M$_\odot {\rm yr}^{-1}$, while {\tt bursty-sn} exhibits a turnover from inflow dominated- to net mass outflow dominated galaxies, reaching outflow rates of 1-10~M$_\odot {\rm yr}^{-1}$ at $L_{\rm [CII]}>10^7$~L$_\odot$. This indicates that [C II] traces a very wide variety of gas flows around galaxies at high redshifts, as had been indicated in Figure~\ref{fig:vradall}. The scatter in the net mass flow rate is small for {\tt smooth-sn} at all luminosities (0.1 dex), while for {\tt bursty-sn} it increases with increasing luminosity (> 1 dex at $L_{\rm [CII]}\geq10^7$~L$_\odot$). 
Inset histograms show the distribution of the net flow rates for the three models for galaxies at $L_{\mathrm{[CII]}} > 10^{7}$~L$_\odot$. {\tt bursty-sn} shows a bimodal distribution, indicating two regimes of gas flows traced by [C II], with $\approx47\%$ of galaxies that are outflow dominated. Meanwhile, {\tt smooth-sn} and {\tt hyper-sn} show continuous distributions that are skewed toward a net inflow dominated regime, with $83\%$ and $80\%$ of galaxies that are inflow dominated,respectively.
We note that even in inflowing galaxies, only about 10$\%$ of the inflowing mass originates from metal-free cosmic filaments (i.e. not contributing to the [CII] luminosity), while most of it comes from recycled gas, sloshed material from past mergers and galactic fountain or companions. 

Figure~\ref{fig:mnetvsLCII} suggests a scenario in which [C II] emitting galaxies are actively driving gaseous outflows (albeit not bright in [C II]; see Figure~\ref{fig:vradall}), yet the overall baryon cycle remains dominated by inflow. Although stellar feedback processes launch substantial outflows, as indicated by the positive correlation between $\dot{M}_{\mathrm{out}}$ and $L_{\mathrm{[CII]}}$, the net mass flux is typically negative, implying that galaxy growth is sustained in this regime.

\begin{figure*}
    \centering
    \includegraphics[width=0.99\linewidth]{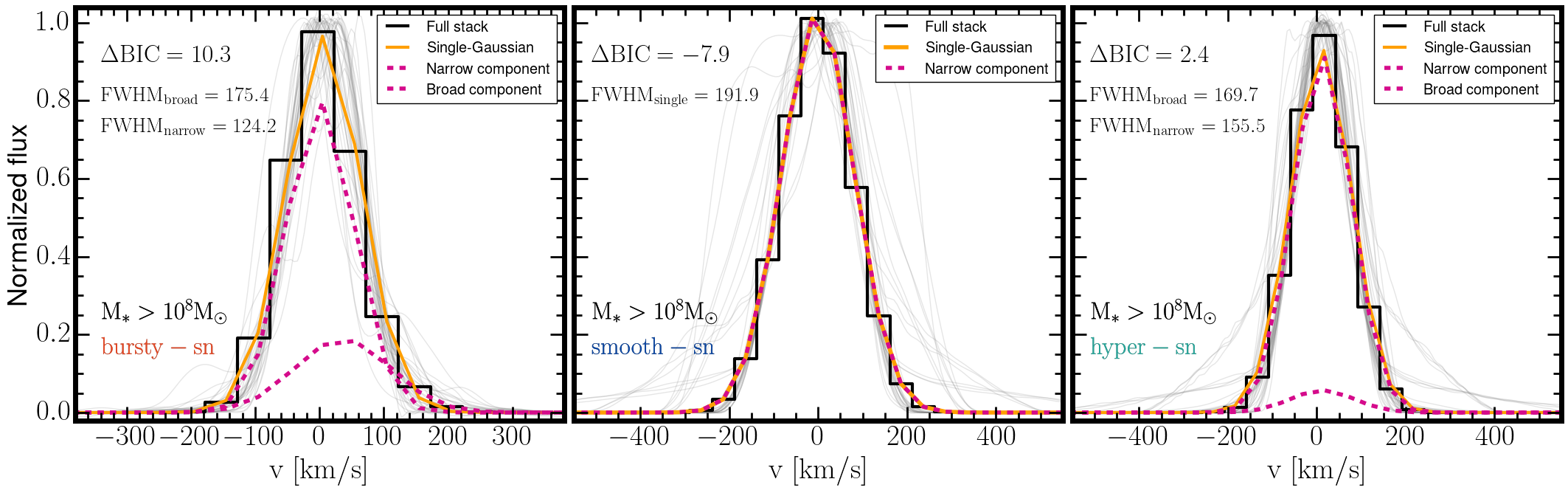}
    \caption{Variance-weighted stacks of the [C II] spectra for all galaxies with $M_* > 10^8 \rm M_\odot$ at $z = 5$–$10$ for the {\tt bursty-sn} (left), {\tt smooth-sn} (middle), and {\tt hyper-sn} (right) feedback models. The stacked [C II] spectra are shown in black, with the best-fitting single-Gaussian models overlaid in orange. The purple dashed curves show the corresponding two-Gaussian fits (narrow + broad components). Grey lines show all individual unscaled spectra from galaxies in each model. Each panel lists the difference in Bayesian Information Criterion values, $\Delta{\rm BIC}$, between the double- and single-Gaussian models. The {\tt bursty-sn} model exhibits strong preference for a two-Gaussian fit ($\Delta{\rm BIC} > 10$), indicating clear evidence for an outflow component. The {\tt smooth-sn} spectra are best described by a single Gaussian ($\Delta{\rm BIC} < 0$), consistent with no detectable outflow, while the {\tt hyper-sn} case indicates only weak evidence for a broad outflow component ($\Delta{\rm BIC} > 0$). FWHM in each panel shows estimated quantities for single or double component Gaussians respectively. }
    \label{fig:specstackall}
\end{figure*}

\begin{figure}
    \centering
    \includegraphics[width=0.96\linewidth]{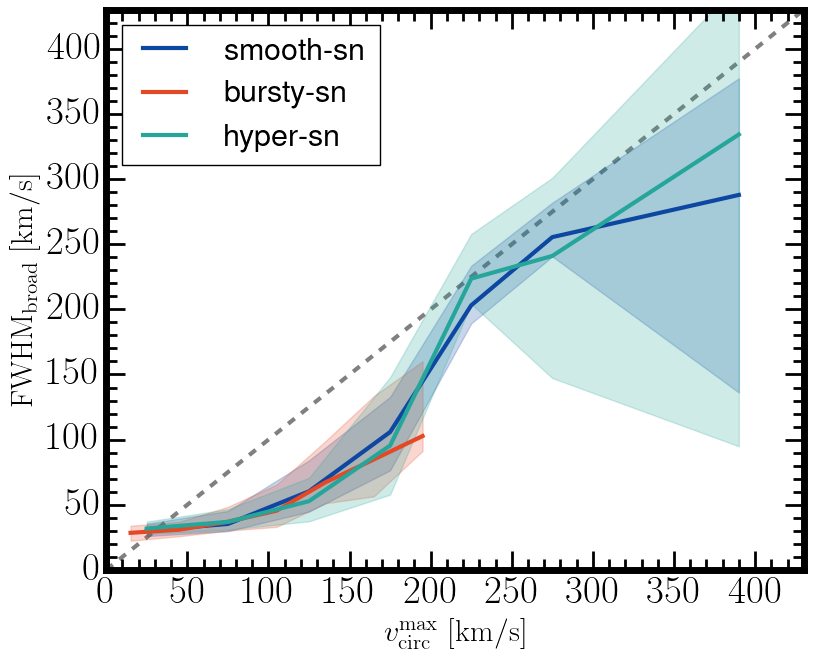}
    \caption{Median FWHM of the broad component of the [C II] line as a function of the maximum circular velocity of the halo calculated within the effective radius of the [C II] emission. Shaded regions represent 16th and 84th percentiles. Grey dashed lines shows the 1:1 line to guide the eye.}
    \label{fig:FWHMvcirc}
\end{figure}
\subsubsection{Outflow velocity}
\label{subsec:vout}
Here we turn to another fundamental diagnostic beyond mass flow rates: the outflow velocity $v_{\rm out}$. As for the mass fluxes, outflow velocities are measured at the effective radius of the [C II] emission. While galactic outflows exhibit a broad and complex velocity distribution, we characterise it using a single representative value, $v_{\rm out}$, defined as the mass outflow rate-weighted 75th percentile of the radial velocity distribution \citep{Nelson_2019}. We estimate outflow velocities for three different tracers: the gas associated with [C II] emission measured directly from the simulations ($v_{\rm out,[CII]}$), the total gas outflow ($v_{\rm out,gas}$) measured from the simulations, and velocities inferred kinematically from synthetic [C II] spectra ($v_{\rm out,obs}$), as described in Section~\ref{subsec:CIIdata}. The spectroscopic estimates assume that a broad [C II] component traces an outflow and can therefore be used to infer an outflow velocity (see Section~\ref{subsec:broadoutflow}).

Figure~\ref{fig:voutvssigmaSFR} shows the outflow velocities as a function of galaxy [C II] luminosity. In the top panel, we compare the outflow velocities of [C II]-bright gas derived directly from the simulations ($v_{\rm out,[CII]}$; dotted lines) with those inferred from the [C II] spectra ($v_{\rm out,obs}$; solid lines). For all three feedback models, both velocity measures increase with $L_{\rm [CII]}$, such that more [C II]-luminous galaxies exhibit larger median outflow velocities. At low luminosities ($L_{\rm [CII]}\sim10^{6{-}7}$~L$_\odot$), typical values of $v_{\rm out,obs}$ are $\sim50{-}60~\mathrm{km\,s^{-1}}$, increasing to $\approx 200{-}300~\mathrm{km\,s^{-1}}$ by $L_{\rm [CII]}\sim10^{9}$~L$_\odot$. In contrast, the corresponding cold gas velocities $v_{\rm out,[CII]}$ are systematically lower, ranging from $\sim20{-}30~\mathrm{km\,s^{-1}}$ at low luminosities to $\sim50{-}100~\mathrm{km\,s^{-1}}$ at the high-luminosity end. The discrepancy between $v_{\rm out,[CII]}$ and $v_{\rm out,obs}$ depends on the feedback model, being largest for the \texttt{smooth-sn} model, followed by \texttt{hyper-sn}, with \texttt{bursty-sn} exhibiting the smallest offset. Across all luminosities and feedback models, the inferred outflow velocities from [C II] spectra exceed the cold-gas velocities measured in the simulations by factors of $\sim2{-}5$.

In the bottom panel of Figure~\ref{fig:voutvssigmaSFR}, we compare the total gas outflow velocities measured directly from the simulations ($v_{\rm out,gas}$; dashed lines) with $v_{\rm out,obs}$ (solid lines). The \texttt{smooth-sn} model consistently produces the lowest total gas outflow velocities, with $v_{\rm out,gas}\lesssim300~\mathrm{km\,s^{-1}}$ even at the highest luminosities. In contrast, the \texttt{bursty-sn} model shows a continued increase in $v_{\rm out,gas}$ with luminosity, reaching $\sim600~\mathrm{km\,s^{-1}}$ at the bright end. The \texttt{hyper-sn} model yields the highest total gas outflow velocities at intermediate luminosities ($v_{\rm out,gas}\sim300{-}500~\mathrm{km\,s^{-1}}$ up to $L_{\rm [CII]}\sim10^{7.5}$~L$_\odot$), followed by a sharp decline at the highest luminosities to $v_{\rm out,gas}\sim300~\mathrm{km\,s^{-1}}$.
At fixed $L_{\rm [CII]}$, the velocities inferred from [C II] spectra are typically a factor of $\sim3{-}10$ lower than the total gas outflow velocities, underscoring that [C II] based measurements trace only a subset of the multiphase outflow.

Overall, Figure~\ref{fig:voutvssigmaSFR} shows that derived outflow properties are sensitive to gas phase and how an outflow is identified and quantified. In particular, outflow velocities inferred from [C II] spectra do not match the full range of cold-gas or total-gas kinematics predicted directly in the simulations (see also Appendix C in \citealt{MartinAlvarez2026}). At $L_{\rm [CII]} \lesssim 10^{6}$~L$_\odot$, the differences in cold and total gas outflow velocities are smallest across the three feedback models. At higher luminosities, however, the simulated gas kinematics diverge significantly between models, while the [C II]–inferred outflow velocities ($v_{\rm out,[CII]}$) remain comparable. At fixed luminosity, estimates based on [C II] spectra systematically yield velocities higher than those measured for the cold gas and lower than those for total gas. This discrepancy reflects the physical limitation of using [C II] as a direct tracer of outflows, as it primarily probes a mix of dense, cold phase and fails to capture the wider multi-phase nature of outflows and their phase-dependent energetics.

\begin{figure*}
    \centering
    \includegraphics[width=0.99\linewidth]{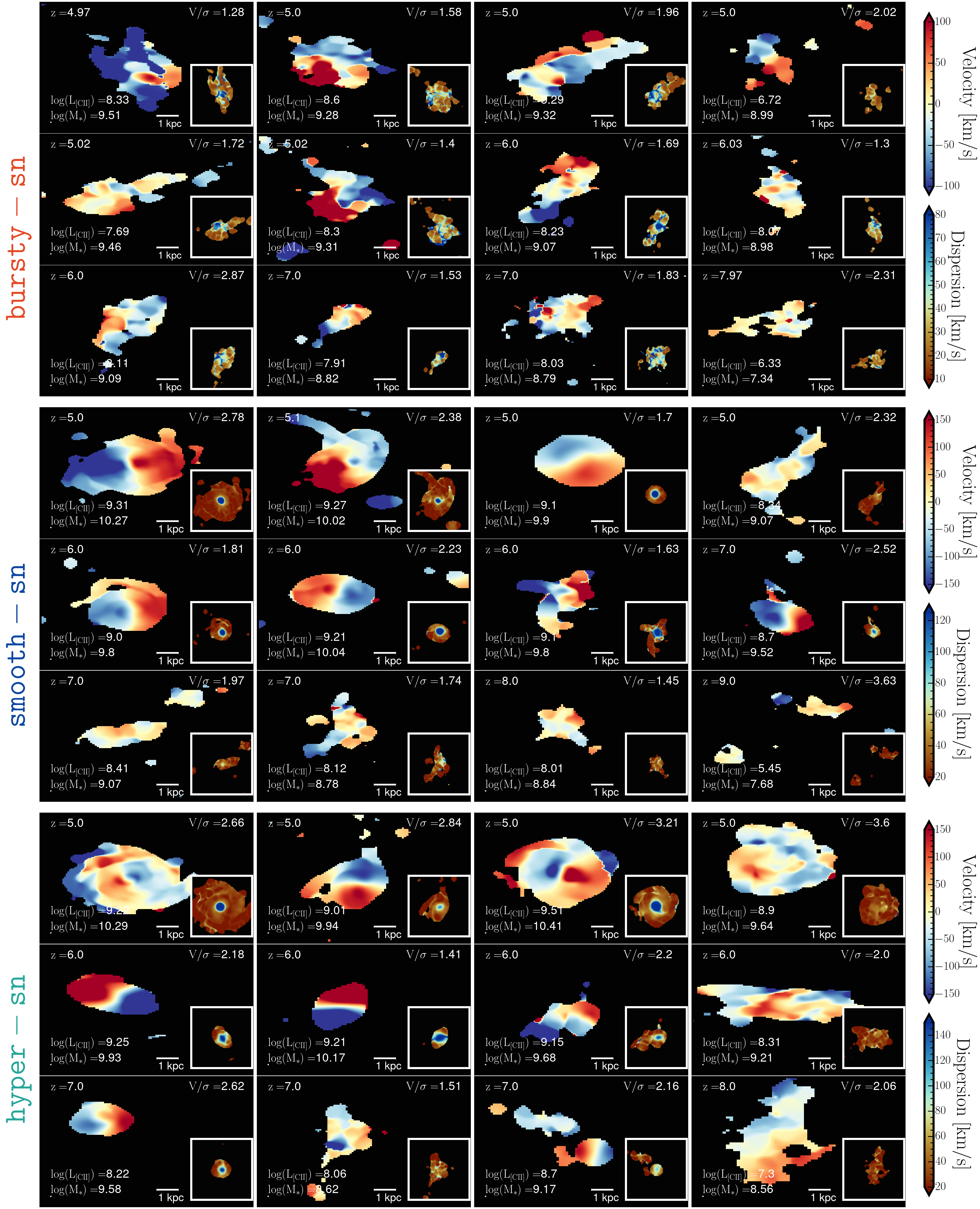}
    \caption{Velocity and velocity-dispersion maps (insets) for {\tt SPICE} galaxies in the three feedback models over $z = 5-8$. The [C II] kinematics exhibit strong sensitivity to the underlying feedback. The {\tt bursty-sn} model produces predominantly irregular and dispersion-dominated systems. In contrast, the {\tt smooth-sn} and {\tt hyper-sn} models forms dynamically cold, rotation-supported disks as early as $z \sim 7$, yielding ordered velocity structures. Overall, we see a large diversity in kinematics traced by [C II].}
    \label{fig:Kinematics}
\end{figure*}

\subsubsection{Feedback energetics}
\label{subsec:feedbackenergetics}

Using mass outflow rate and velocity estimates, one can further estimate the outflow kinetic power at the [C II] emission radius. Estimates of outflow kinetic power are often used as proxy for the efficiency of stellar feedback and constrain the baryon cycle in high-redshift galaxies. A common approach in the literature \citep{Hailey_Dunsheath_2010, 2020Ginolfi, Herrera_Camus_2021} is to evaluate the mass-loading factor, $\eta = \dot{M}_{\rm out}/{\rm SFR}$, as a measure of feedback efficiency\footnote{The normalization of $\eta$ is very sensitive to the choice of gas phase and radius of calculation, however, the qualitative trends shown do not vary.}. 

Observationally, $\eta$ is inferred by estimating the mass of the outflowing atomic gas as $
    M_{\rm out}^{\rm atom} = 0.7 \biggl( \frac{0.7 L_{\rm [CII]}}{L_\odot} \biggr)  \biggl( \frac{1.4\times10^{-4}}{X_C^+} \biggr) \biggl( \frac{1 + 2e^{-91/T_K} + n/n_{\rm crit}}{ 2e^{-91/T_K}} \biggr)$ \, ,
where $L_{\rm [CII]}$ is the luminosity of the outflowing region, $X^+_C = 1.4 \times10^{-4}$ is the carbon abundance, $T_K = 100 \rm K$ is the gas temperature and $n_{\rm crit} = 3\times10^3 \ \rm cm^{-2}$ is the critical density \citep{Hailey_Dunsheath_2010,2020Ginolfi,birkin2025alma}. Together with the observed outflow velocity ($v_{\rm out, obs}$) and the effective [C II] emission radius ($R_{\rm [C II]}$), one can compute the atomic mass outflow rate as $\dot{M}_{\rm out}=v_{\rm out, obs} M_{\rm out}^{\rm atom}/R_{\rm [C II]}$. The resulting $\dot{M}_{\rm out}$, when compared to the SFR, provides an estimate of the mass-loading factor $\eta$.

To assess the reliability of using [C II] as a tracer of feedback efficiency, we compute $\eta$ both directly from simulations ($\eta_{\rm gas}$) and using the observationally motivated approaches ($\eta_{\rm obs, [C II]}$) described above for all three feedback models. Figure~\ref{fig:etavsMstar} shows the median estimated $\eta_{\rm gas}$ (dashed lines) and $\eta_{\rm obs, [C II]}$ (solid lines) as a function of stellar mass for [C II] emitting galaxies. Across all feedback models, $\eta_{\rm gas}$ shows a decline in mass-loading factors with increasing stellar mass, likely due to outflow suppression in the deeper gravitational potentials of the massive galaxies \citep{Dekel986}. 
Predictions from {\tt SPICE} are consistent with previous simulations such as FIRE-2, as shown with the gray dashed lines \citep{2021Pandya}. 
In contrast, the [C II] spectra-derived $\eta_{\rm obs, [CII]}$ displays contrasting behaviour. It increases monotonically with stellar mass, as the outflowing mass scales with $L_{\rm [CII]}$. We see a sharp disagreement between $\eta_{\rm gas}$ and $\eta_{\rm obs, [C II]}$, particularly at $M_* < 5\times10^8$~M$_\odot$, where the mass loading differs by a factor of $\sim100-1000$. Interestingly, estimates of $\eta_{\rm gas}$ and $\eta_{\rm obs, [C II]}$ for the {\tt smooth-sn} and {\tt hyper-sn} converge at $M_* > 10^9$~M$_\odot$, and show good agreement up to $\approx10^{10}$~M$_\odot$. At $M_* \gtrsim 10^{10.5}$~M$_\odot$, which cannot be probed in {\tt SPICE}, AGN feedback may create an upturn in $\eta_{\rm gas}$  \citep[][]{Dubois2013, Costa2014, Costa2015, Biernacki2017,Nelson_2019}. {\tt bursty-sn} shows a lack of convergence between $\eta_{\rm gas}$ and $\eta_{\rm obs, [C II]}$ even at the highest masses. This offset between the scaling-relation-based estimates and simulated predictions suggests a fundamental limitation of [C II] as a tracer of feedback. Since [C II] does not directly probe the feedback-driven outflowing phase (Figure~\ref{fig:vradall}), but instead traces colder, dynamically slower gas, it leads to systematically underestimated values of $\eta_{\rm gas}$.

\subsubsection{Broad wings in [C II]}
\label{subsec:broadoutflow}
Typically, the presence of a broad component in [C II] spectra is regarded as evidence for an outflowing component of gas \citep{Herrera_Camus_2021}. Additionally, with spatially resolved observations, one can map the outflowing region (e.g middle right panel in Figure~\ref{fig:UVCIIcontours}). However, in observations of star-forming galaxies, detecting the broad component remains a challenge, and stacking is used to boost S/N and find evidence for outflows \citep{2020Ginolfi,birkin2025alma}. 
In this section, we mimic this approach and test whether an outflow signal emerges in stacks from the three feedback models in {\tt SPICE}.  

In Figure~\ref{fig:specstackall}, we show the variance-weighted stacks for the three feedback models for all galaxies with $L_{\rm [C II]} > 10^8$~M$_\odot$, obtained by following the stacking procedure of \citet[][see section 3.3]{birkin2025alma}. To summarize, we scale the line width of each spectrum to the median of the sample, adjust the flux density by the peak value, and then stack by weighting with the variance. This method ensures that the line profiles are studied consistently. Additionally, as in \citet{birkin2025alma}, we use the BIC (see section~\ref{subsec:CIIdata}) to determine if a broad component is required to model the stacked emission line. 

{\tt bursty-sn} favors the presence of an outflowing component as indicated by $\Delta \rm BIC > 10$ (where $\Delta \rm BIC = BIC_{\rm single}-BIC_{\rm double}$), with an estimated outflow velocity of $ 326.3 \ \rm km/s$. Meanwhile, {\tt smooth-sn} is best described by a single gaussian fit with $\Delta \rm BIC < 0$ indicating that a second (broad) component is not statistically required. Finally, {\tt hyper-sn} shows weak evidence for an outflowing broad component with $\Delta \rm BIC = 2.4$, with an inferred outflow velocity of $308 \ \rm km/s$. The outflow trends derived from stacking are in agreement with overall trends shown in Figure~\ref{fig:mnetvsLCII}, as {\tt bursty-sn} is outflow dominated, {\tt hyper-sn} exhibits a small fraction of galaxies that are outflow dominated, while {\tt smooth-sn} is generally lacking evidence for strong outflows. 

Finally, despite {\tt bursty-sn} demonstrating strongest evidence for outflows in stacks and through derived mass flow rates, we see the broadest [C II] lines in {\tt smooth-sn} and {\tt hyper-sn} (as demonstrated in Figure~\ref{fig:UVCIIcontours} and grey individual spectra shown in Figure~\ref{fig:specstackall}). Therefore, {\tt SPICE} galaxies suggest a dynamical origin of broad lines in [C II] as opposed to feedback driven outflows (also suggested by Figures~\ref{fig:voutvssigmaSFR},~\ref{fig:etavsMstar}). To confirm this, in Figure~\ref{fig:FWHMvcirc} we show the relation between the FWHM of the broad component of [C II] as a function of maximum circular velocity of the halo. The circular velocity is derived as $v_{\rm circ} = \sqrt{GM(<r) /r}$ where $G$ is the Gravitational constant, $M(<r)$ is the total mass enclosed with radius $r$. We calculate this for shells with maximum radius equal to the effective radius of [C II] emission and $v_{\rm circ}^{\rm max}$ is the maximum value within this radius. This quantity serves a proxy to understand the depth of the gravitational potential and dynamical movement of gas due to gravity.

The FWHM of the broad [C II] component increases with the maximum circular velocity across all feedback models, indicating that the line width primarily traces the depth of the gravitational potential rather than outflow strength. At fixed $v_{\rm circ}^{\rm max}$, the {\tt bursty-sn} model systematically exhibits lower FWHM despite showing clear signatures of outflows in stacked velocity profiles, consistent with coherent, bulk radial motions that do not strongly broaden the line. In contrast, the {\tt smooth-sn} and {\tt hyper-sn} models produce significantly larger FWHM, particularly at high $v_{\rm circ}^{\rm max}$, despite lacking strong outflow signatures (see Figure~\ref{fig:specstackall}). These systems are characterized by more centrally concentrated morphologies (see \citet{bhagwat2024}), suggesting that the large line widths arise from dispersion-dominated, gravitationally supported gas rather than feedback-driven winds. Together, these results demonstrate that broad [C II] line widths do not uniquely trace outflowing gas, but are strongly influenced by the underlying potential and galaxy structure.

\subsection{Morpho-kinematic classification}
\label{subsec:morphokinematic}

In the previous sections,  we discuss [C II] as a tracer of gas flows. Here we quantify the kinematics of [C II] emitters. Figure~\ref{fig:Kinematics} shows the velocity and velocity dispersion (inset) maps of the brightest [C II] emitters for the three feedback models in the range $z=5-8$. We note a strong diversity in the kinematics of [C II] emitters in {\tt SPICE}. We quantify these trends in two ways, as described in section~\ref{subsec:CIIdata}: 3D $V/\sigma$ estimated using rotation curves from the simulations, and an observational style morpho-kinematic classification to identify rotators and dispersion dominated galaxies following the method entailed in \citet{Lilian2025}.

We use the method described in \citet{Lilian2025} to classify {\tt SPICE} galaxies in disk- and non disk-like systems based on a disk score. Briefly, each galaxy can be given a maximum score of 8, where 1 point is obtained if $V_{\rm obs}/2\sigma_{\rm int}$ > 0.4, 2 points if the P-V diagram is devoid of clumps and disconnected structures,
2 points if the centroid of emission across velocity channels follows a monotonous locus, 2 points if the velocity and dispersion asymmetry ($v_{\rm asym}$ and $\sigma_{\rm asym}$) are such that $K_{\rm asym} = (v_{\rm asym} + \sigma_{\rm asym})^{1/2} < 0.5 $, and 1 point if there are no companions within 2 kpc of emission centroid. We define a system disk-like if it has a score $\geq5$.

\begin{figure}
    \centering    \includegraphics[width=0.99\linewidth]{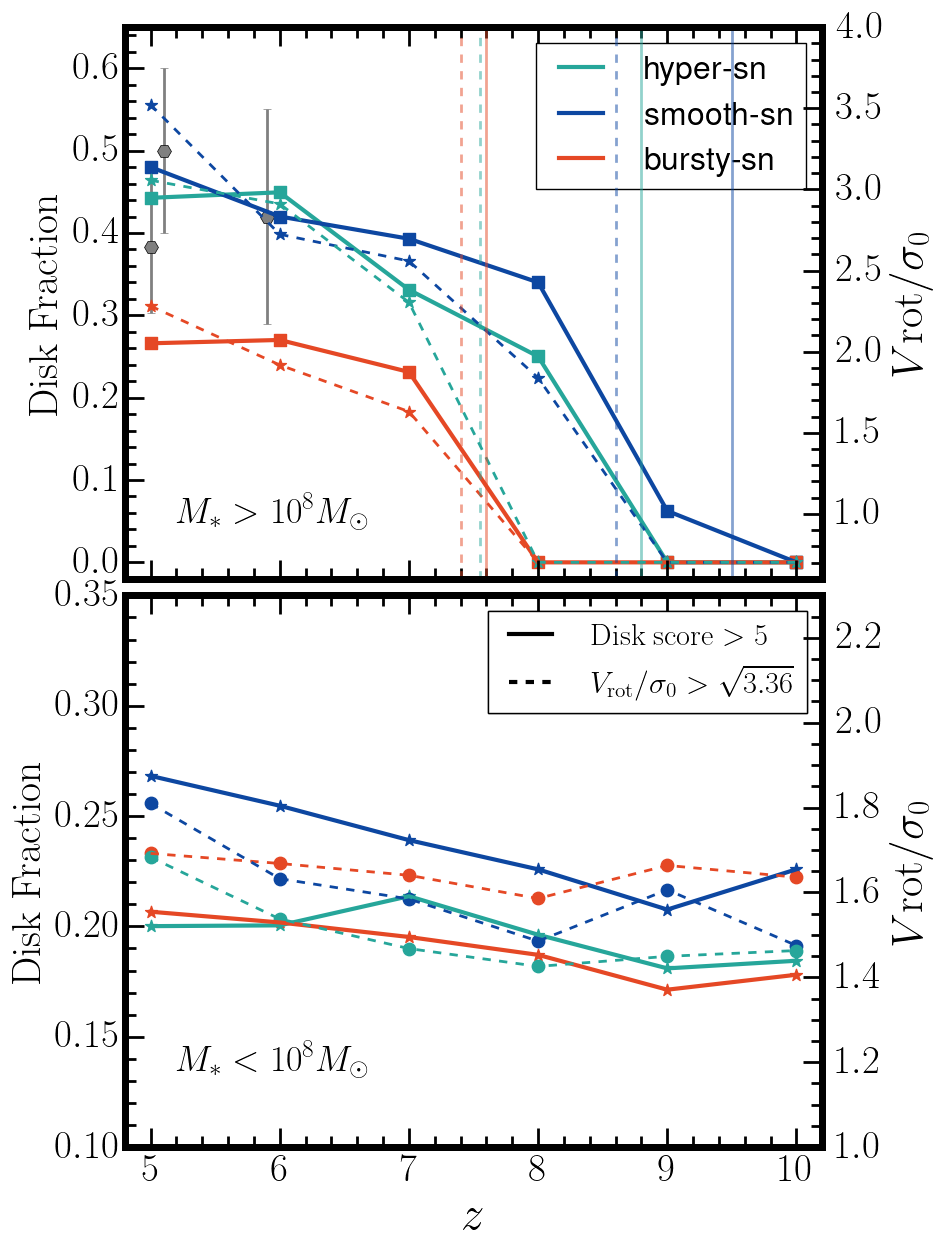}
    \caption{{\it Top:} Redshift evolution of the fraction of galaxies with $M_* > 10^8$~M$_\odot$ classified as disks using a morpho-kinematic classification (solid lines) and from 3D rotation curves described in section~\ref{sec:MethodsCII} (dashed) for {\tt bursty-sn} (red), {\tt smooth-sn} (blue) and {\tt hyper-sn} (turquoise). {\it Bottom:} As in the top panel but for galaxies with $M_* < 10^8$~M$_\odot$. Grey data points show disk fraction estimates from \citet{2023Ferreira} and \citet{Lilian2025}. Vertical solid (dashed) lines show the timing of the emergence of the first disk galaxy when using morpho-kinematic classification (3D rotation curves) for the three feedback models. }
    \label{fig:DiskFrac}
\end{figure}

Using this metric, in Figure~\ref{fig:DiskFrac} we show the redshift evolution of disk fractions (left axis) of [C II] emitters. The top panel shows the redshift evolution of disk fractions for $M_* > 10^8 \rm M_\odot$ as estimated using morpho-kinematic classification (solid lines) and from 3D rotation curves described in the previous section (dashed). Disk fractions estimated from morpho-kinematic classification  for {\tt smooth-sn} rise from $\approx6\%$ at $z=9$ to $\approx47\%$ at $z=5$, demonstrating early settling of cold disks. For {\tt hyper-sn} disks emerge later, at $z<9$, with disk fractions rapidly rising to $\approx 28\%$ at $z=8$ and to $\approx43\%$ at $z=5$. Finally, {\tt bursty-sn} shows the latest emergent disks ($z < 8$) and the lowest disk fractions of $\approx 23\%$ at $z=7$, rising to $28\%$ at $z=5$. The bottom panel of Figure~\ref{fig:DiskFrac} is the same as the top one, but for $M_* < 10^8 M_\odot$. In this case we see that the disk fractions for {\tt hyper-sn} and {\tt bursty-sn} increase almost negligibly between $z=10$ and 5, with a value of $\approx 20-23\%$. {\tt smooth-sn}, instead, shows a mild rise, with disk fractions of $\approx 22-27\%$. At $z\sim5{-}6$ the disk fractions predicted by \texttt{smooth-sn} and \texttt{hyper-sn} are broadly consistent with observational constraints from ALMA-CRISTAL \citep{Lilian2025} and optical surveys \citep{2023Ferreira}, whereas \texttt{bursty-sn} underpredicts the observed fractions by $\sim0.1{-}0.2$ dex.
Figure~\ref{fig:DiskFrac} shows the evolution of the median $V_{\mathrm{rot}}/\sigma_0$ (right axis; dashed lines) calculated from 3D rotation curves as a function of redshift. For $M_* > 10^8$~M$_\odot$, all models exhibit increasing rotational support, with disk-like systems ($V_{\mathrm{rot}}/\sigma_0 > \sqrt{3.36}$; \citealt{Schreiber_2018}) forming earliest in {\tt smooth-sn} ($z\simeq8$), followed by {\tt hyper-sn} ($z\simeq7.5$), and latest in {\tt bursty-sn} ($z\simeq5$). Similarly to the previous method, galaxies with $M_* < 10^8$~M$_\odot$ remain below this threshold down to $z=5$, although $V_{\mathrm{rot}}/\sigma_0 > 1$ in all models indicates the presence of ordered rotation in the cold gas.

The observational style morpho-kinematic approach and theoretical rotation curves show generally consistent trends with redshift. 
{\tt hyper-sn} and {\tt bursty-sn} show excellent agreement between the disk fractions from the two methods curves at all stellar masses. For {\tt smooth-sn} the morpho-kinematic classification predicts much higher disk fractions (by $\approx 0.5 \rm dex$) for both mass cuts. Only at $z\sim5$ the 3D rotation curves predicts a disk fraction of $\approx0.7 \rm ~dex$ higher than the morpho-kinematic classification for {\tt smooth-sn}.
Importantly, the morpho-kinematic classification predicts an earlier emergence of disk-like galaxies as compared to 3D rotation curves for $M_* > 10^8$~M$_\odot$ for all models. The difference is minimal for {\tt bursty-sn}, where the latter predicts the emergence of disks at $z\approx7.2$, while the former at $7.6$. The differences are large for {\tt smooth-sn} and {\tt hyper-sn}, where morpho-kinematic estimates show first disk formation at $z\approx9.5$ and $8.7$, respectively, while rotation curves predict $z\approx8.8$ and $7.4$, respectively. Therefore, while the methods generally agree on disk fractions, the timing of onset of disk formation can be differ by $\Delta z \approx 0.4 - 1.2$.  

Figure~\ref{fig:DiskFrac} demonstrates that the dynamical state of high-redshift galaxies depends on both stellar mass growth and the nature of supernova feedback. Bursty feedback maintains higher dispersion, suppresses rotational support and delays disk-formation, whereas smoother feedback enables the early emergence of disk-like systems. The evolution of $V_{\rm rot}/\sigma_0$ and disk fraction, therefore, traces the balance between feedback-driven dispersion and gravitational settling over cosmic time, predominantly for galaxies with $M_* > 10^8$M$_\odot$. Additionally, differences between classification methods further show that inferred disk fractions depend sensitively on the adopted kinematic definition.

\section{Discussion}
\label{sec:discussion}

\subsection{[C II] as a probe of a feedback-regulated ISM}

A tight correlation between $L_{\rm [CII]}$ and SFR has been established from the local Universe to $z>6$ \citep{DeLooze2014FIRSFR,Schaerer_2020}. Recent ALMA programs targeting reionisation-era galaxies report broadly linear scaling with increased scatter toward low mass and high specific SFR \citep{Carniani_2020}. The {\tt SPICE} simulations reproduce this global trend in the range $z=10-5$ for all feedback prescriptions, confirming that [C II] remains a reliable tracer of star formation in rapidly assembling systems.

However, a bursty feedback systematically suppresses $L_{\rm [CII]}$ at fixed SFR and increases the intrinsic scatter with respect to the {\tt smooth-sn} and {\tt hyper-sn} feedback models. This behaviour is consistent with observed [C II] deficit in compact starbursts and extreme emission-line galaxies \citep{Smit2018,Carniani_2020}. The physical interpretation provided by {\tt SPICE} is that bursty feedback reduces the volume filling fraction of neutral gas ($f_{\rm HI}$), thereby shrinking the reservoir capable of producing [C II] emission even when the SFR remains high but spatially compact bursts. This mechanism differs from metallicity-driven suppression  scenarios \citep{Bisbas_2022} since {\tt SPICE} galaxies do not reach saturation threshold, and instead links the [C II] luminosity directly to feedback-regulated ISM phase structure.

JWST has now revealed highly compact star-forming regions with intense radiation fields at $z\gtrsim6$ \citep{hunt2025,Tang2025}. In such environments, bursty feedback and strong radiative feedback can evacuate channels in neutral gas (decreasing the PDR fraction) on short timescales. {\tt SPICE} suggests that this restructuring of the ISM rather than suppressed star formation efficiency can naturally produce the increased dispersion and occasional [C II] suppression observed in these systems.

The observed suppression or non-detections in the $L_{\rm [CII]}$–SFR relation at high redshift may encode information about feedback duty cycles. Combining statistical ALMA [C II] measurements with JWST diagnostics (e.g., [O III]/H$\beta$ ratios) could break the degeneracy in feedback and constrain the temporal variability of star formation.

\subsection{Spatial Structure of [C II]: Gas Redistribution or Distributed Star Formation}

High-resolution ALMA imaging has consistently shown that [C II] emission extends beyond the rest-frame UV continuum by factors of $\sim2$–5 \citep{Fujimoto_2019,Fudamoto_2022}. The physical interpretation remains debated, as extended [C II] could trace diffuse star formation, cold accretion streams, or feedback-driven gas redistribution.

{\tt SPICE} reproduces the observed offset, with $R_{\rm [CII]}/R_{\rm UV}\sim2$–4 across all models. However, the feedback model determines how this extended emission arises. In {\tt smooth-sn}, [C II] is relatively compact and correlated with stellar structure, while in {\tt bursty-sn}, the emission is spatially extended and morphologically irregular owing to a non-uniformly reduction of $f_{\rm HI}$. The increase in absolute [C II] area with luminosity, combined with a decrease in area relative to halo size for all three feedback models, indicates that luminous systems concentrate neutral gas centrally, while low-luminosity galaxies retain extended, diffuse structures.

This behavior is consistent with recent observational trends showing larger [C II] half-light radii in clumpy or irregular galaxies \citep{2024Posses}. Importantly, {\tt SPICE} demonstrates that extended [C II] does not require spatially extended star formation. Instead, bursty feedback redistributes gas to larger radii while preventing central gas concentrations, increasing area emitting [C II] without increasing the SFR surface density.
Future JWST imaging, combined with resolved ALMA maps, can directly test whether extended [C II] correlates with UV clumpiness, Gini index, or stellar compactness. A strong correlation would support a feedback-driven redistribution scenario rather than in-situ extended star formation.

\subsection{On the Interpretation of [C II] as a Tracer of Multiphase Gas Flows}

[C II] emission is widely used to trace gas flows in high-redshift galaxies, however, it remains unclear if the gas dynamics in [C II] emitters are dominated by galactic winds. In {\tt SPICE} outflows are ubiquitous at $R_{\rm [CII]}$ across all feedback prescriptions (see Figure~\ref{fig:mnetvsLCII}), and as demonstrated in section~\ref{subsec:ciiflows} are a mix of feedback driven outflowing and gravitational motion. However, galaxies remain overall net inflow-dominated at these radii. This supports a picture in which [C II] emitters are governed by gas cycling in a galactic fountain \citep{2009Dekel}, rather than sustained large-scale gas expulsion. Observational evidence for continued accretion, such as filamentary inflows in $z\sim6$ systems \citep{2016Micha,Jones_2017Fil}, is broadly consistent with this interpretation. 

Gas flows imprint signatures such as the presence of a broad component in spectral profiles. Broad [C II] components have therefore frequently been interpreted as signatures of galactic outflows and used to estimate outflow velocities \citep{2018Gallerani,2019Sugahara,2020Ginolfi}. However, predictions from {\tt SPICE} suggest that outflow velocities inferred from [C II] do not reliably trace the multiphase wind and instead misrepresent the properties of the cold outflowing gas.
Measurements based on [C II] spectra are sensitive to the complex kinematic structure of the emitting gas. As previously shown (section~\ref{subsec:broadoutflow}), the [C II] kinematics can be set by inflows as opposed to outflowing [C II] bright material (see Figure~\ref{fig:voutvssigmaSFR}) which can lead to an overestimation of its bulk radial velocity when interpreted as an outflow speed. 
At the same time, since outflows are energetically dominated by hot gas (e.g. \citealt{Hu2019,Fielding_2022}), [C II] systematically underestimates the velocities of the fastest components of the multiphase wind (see \citealt{Cicone_2018,Fluetsch2019}), likely because it does not trace feedback-driven winds.

These velocity biases propagate directly into estimates of feedback energetics such as mass-loading factors. In {\tt SPICE}, theoretical mass-loading factors ($\eta_{\rm gas}$) measured at $R_{\rm [CII]}$ decline with increasing stellar mass, consistent with expectations for energy-driven winds \citep{Nelson_2019, 2021Pandya}. However, mass-loading factors inferred from [C II] spectra ($\eta_{\rm obs, [CII]}$) systematically underestimate the theoretical values. This arises because inferred mass outflow rates depend on the assumed velocity, and wind kinetic energy scales as $v^2$--$v^3$ depending on the adopted geometry. Missing the high-velocity, hot phase therefore leads to a systematic suppression of the inferred mass-loading factors.

Finally, we find that the width of the broad [C II] component is not uniquely linked to outflow strength. While stacked spectra reveal a statistically significant broad component (Figure~\ref{fig:specstackall}) in the {\tt bursty-sn} feedback model, consistent with the presence of outflowing gas \citep{2019Sugahara}, the largest FWHM values are found in the {\tt smooth-sn} and {\tt hyper-sn} feedback models, which show weaker or no clear outflow signatures. Instead, we show that FWHM correlates strongly with the circular velocity of the halo, indicating that it primarily traces the depth of the gravitational potential. This demonstrates that broad [C II] lines can therefore arise from both feedback-driven motions and gravitationally induced kinematics, and do not uniquely trace outflows.

Taken together, these results imply that [C II]-based outflow measurements provide a lower bound on total wind energetics, particularly in dwarf galaxies where the hot phase dominates the kinetic energy budget. At the same time, the net inflow-dominated nature of galaxies at $R_{\rm [CII]}$ is consistent with pseudo-equilibrium ``bathtub'' models \citep{2009Dekel,2013Lilly} which predict sustained accretion even in the presence of active ejective feedback. Recovering the full multiphase structure of galactic winds will therefore require combining ALMA [C II] kinematics with complementary tracers, such as rest-frame optical emission lines with {\tt JWST} or future constraints on the hot phase from X-ray observations.

\subsection{Disk Formation and Kinematic Settling}

The {\tt SPICE} simulations show that the emergence of rotational support is strongly regulated by the mode of feedback (i.e bursty vs smooth, see~\citealt{bhagwat2024}), with important implications for the interpretation of high-redshift galaxy kinematics. While all models exhibit increasing $V_{\rm rot}/\sigma_0$ and disk fractions from $z=10$ to $z=5$, smoother feedback promotes earlier development of rotationally supported systems, whereas bursty feedback maintains elevated dispersions and delays this transition. This behavior is broadly consistent with previous zoom-in simulations in which bursty star formation cycles sustain dynamically hot gas and delay disk assembly \citep{Gurvich_2022,Hopkins_2023,Yu_2023,2025McClymont}. In contrast to the classical picture of disc formation, in which gas settles into dynamically cold configurations after angular momentum conservation \citep{Fall1980,Mo1998}, our results indicate that at high redshift the emergence of rotational structure does not necessarily coincide with dynamically cold kinematics. {\tt SPICE} galaxies develop ordered rotation while retaining substantial velocity dispersion, implying that disc assembly and full dynamical settling are not equivalent processes \citep{Mo2024}.

This distinction is particularly relevant in light of recent observational claims that some galaxies at $z\gtrsim4$--7 already host dynamically cold, ordered discs. High-resolution ALMA observations have identified several [C II] rotators with large inferred $V/\sigma$ \citep{Rizzo2020,Lelli2021,Fraternali2021}, while theoretical work suggests that such systems may arise under favourable conditions but are sensitive to tracer choice and measurement methodology \citep{Kohandel_2020_veldisp, Kohandel_2024_myth, phillips2025}. At the same time, emerging population-level constraints from JWST indicate that although disky systems are present at $z\sim4$--6, galaxies remain significantly more turbulent on average \citep{2025Danhaive}. {\tt SPICE} supports this broader picture shown in \citet{2025Danhaive}: rotationally organized systems emerge by $z\sim6$-5, yet retain significant non-circular motions, suggesting that the coldest observed discs likely represent the upper envelope of disc settling rather than the norm.

A further implication is that the inferred kinematic state depends on both the tracer and the method used to identify discs. While morpho-kinematic classifications and 3D rotation curves yield broadly consistent disc fractions, they differ systematically in the inferred timing of disc formation, with offsets of $\Delta z \sim 0.4$-1.2. Morpho-kinematic methods \citep{Lilian2025} tend to identify discs earlier, as they are sensitive to projected velocity gradients and morphology, whereas rotation-curve-based diagnostics provide a stricter test of whether ordered rotation dominates over random motions. This highlights the importance of methodological consistency when comparing simulations and observations \citep{lee2024codes}.

As shown in previous sections, [C II] emission does not simply trace the bulk rotation of a dynamically cold disc, but is also sensitive to non-rotating motion in the cold gas. In particular, the FWHM of [C II] correlates strongly with the circular velocity of the halo, indicating the observed dispersion is set by the depth of the potential rather than feedback-driven winds alone. Consequently, large [C II] line widths do not necessarily imply the absence of a disc, just as ordered velocity gradients do not necessarily imply a dynamically settled system. These results point toward a more nuanced picture of disc settling at high redshift. While early disc assembly is common, most galaxies remain dynamically hot due to ongoing accretion, gas cycling, and gravitational structure, requiring a clear distinction between the emergence of rotational support and the attainment of dynamically cold configurations.

\subsection{Combining [C II] with multi-wavelength observations}

Large JWST surveys such as JADES, CEERS, GLASS, and UNCOVER have enabled spatially resolved measurements of H$\alpha$ and [O III] in galaxies at $z\gtrsim5$ \citep[e.g.][]{Wang_2022,Curti2023,Trump2023, Tripodi_2024, hsiao2024, Tang2025}. These lines directly probe H~II regions, nebular excitation, ionisation structure, and radiation hardness. At the same time, ALMA surveys including ALPINE and CRISTAL have characterized [C II] luminosities, sizes, and kinematics in statistically meaningful samples \citep{LeFevre2020,2025Herrera-Camus}. 

The {\tt SPICE} results demonstrate that [C II] alone provides an incomplete view of feedback-regulated galaxy evolution at high redshift. Indeed, while [C II] robustly traces the cold ISM and its connection to star formation, it samples only a subset of the multiphase medium and under-populates the energetically dominant ionised and hot outflow components. Interpreting observed [C II] luminosities, spatial extent, and kinematics, therefore, requires complementary constraints on ionised gas phases now accessible with JWST through observations of balmer lines and strong emission lines such as [O III], [O II], [N II] etc. Observational campaigns such as {\tt GA-NIFS} \citep{Parlanti_2025,2025Scholtz} have targeted [C II] emitters with {\tt JWST} to obtain strong constrains on multi-phase outflows using aforementioned emission lines.  

Within this framework, {\tt SPICE} yields several testable multi-wavelength predictions with the goal of finding observable stringent constraints on feedback. Galaxies with bursty feedback as in {\tt bursty-sn} exhibit suppressed $L_{\rm [CII]}$ at fixed SFR and reduced $f_{\rm HI}$ should display elevated nebular excitation and strong [O III]/H$\beta$ ratios (Bhagwat et al. in prep.), similar to the extreme emission-line systems now observed with JWST \citep{Curti2023,Tang2025, hunt2025}. Conversely, compact, rotation-supported systems with weaker feedback as in {\tt smooth-sn} are expected to show more spatial coincidence between [C II] and H$\alpha$, reflecting a more coherent neutral gas reservoir. Kinematically, JWST detections of broad H$\alpha$ or [O III] components \citep{Carniani_2024} provide an independent probe of ionised outflows, enabling direct comparison to the phase-dependent velocity offsets we find between cold gas outflow velocity, spectroscopically inferred outflow velocity, and all gas outflow velocity (see section~\ref{subsec:vout}).

Crucially, the feedback-driven restructuring of the neutral ISM in {\tt bursty-sn} has implications for the escape of ionising radiation. Fragmented neutral media and reduced $f_{\rm HI}$ are expected to facilitate higher escape fractions, whereas compact, rotationally supported systems in {\tt smooth-sn} should attenuate ionising photons more efficiently. Observationally, JWST has begun to identify candidate density-bounded and high-excitation systems potentially associated with elevated escape fractions \citep{Tang2025,Giovinazzo_2026,2026Tripodi}. A combined analysis of [C II] morphology, H$\alpha$ surface brightness, and nebular excitation therefore provides a physically motivated pathway to constraining ionising escape in the reionisation era. A coordinated JWST–ALMA approach thus enables reconstruction of the multiphase ISM: [C II] constrains the neutral reservoir and gas redistribution, H$\alpha$ traces ongoing star formation and ionised kinematics, and [O III] probes excitation and radiation hardness. Embedding these diagnostics within the {\tt SPICE} framework offers a predictive route to linking feedback variability, gas cycling, and ionising photon escape a connection we explore in future work (Bhagwat et al. in prep.).

\section{Conclusions}
\label{sec:conclusionsCII}

In this work, we have used the cosmological RHD simulation suite \textsc{SPICE} \citep{bhagwat2024} to investigate how the temporal structure and energetics of supernova feedback shapes [C II] emission and the baryon cycle in high-redshift galaxies. By systematically varying the supernova feedback models namely: {\tt bursty-sn}, {\tt smooth-sn}, and {\tt hyper-sn}, which span a range of star formation stochasticity from bursty to smooth while keeping all other physical ingredients fixed, we isolate their impact on the neutral ISM, gas redistribution, outflow energetics, and disk assembly. Our main conclusions are as follows:

\begin{enumerate}

\item \textit{The slope of the $L_{\rm [CII]}$--SFR relation does not depend on feedback, but its normalisation and scatter do.}
All three feedback models recover a strong correlation between $L_{\rm [CII]}$ and SFR at all redshifts. However, {\tt bursty-sn} suppresses $L_{\rm [CII]}$ at fixed SFR and increases intrinsic scatter (section~\ref{subsec:luminosity}), particularly below $L_{\rm [CII]}\sim10^7$~L$_\odot$. This suppression coincides with a systematic decline in the neutral gas filling fraction, $f_{\rm HI}$, demonstrating that [C II] luminosity is regulated by the survival and clumping of cold neutral gas.

\item \textit{Neutral gas structure varies systematically with feedback and [C II] luminosity.}  
In {\tt bursty-sn}, $f_{\rm HI}$ decreases with increasing $L_{\rm [CII]}$, whereas {\tt smooth-sn} maintains a more stable neutral fraction across luminosities (section~\ref{subsec:luminosity}). The {\tt hyper-sn} model exhibits an intermediate behaviour. These trends directly link variation in feedback models to ISM phase balance. At fixed SFR, dispersion in $f_{\rm HI}$ modulates the neutral gas reservoir that can emit [C II], driving scatter in $L_{\rm [CII]}$.

\item \textit{[C II] emission is extended relative to the UV component and traces redistributed gas.}  
[C II] radii exceed UV radii by factors of $\sim2$--4 in all models (section~\ref{subsec:radcii}), with the largest offsets in {\tt bursty-sn}. The projected [C II] area increases with luminosity but decreases relative to dark matter halo area, indicating that luminous systems concentrate neutral gas within the inner halo, while lower luminosity systems exhibit more diffuse distributions. Extended [C II] emission therefore reflects feedback-driven gas redistribution rather than spatially extended star formation alone. Satellite galaxies contribute $<0.1\%$ of the total [C II] luminosity.

\item \textit{[C II] velocity distributions reveal a bias toward slow, recycled gas.}  
Radial velocity distributions show that while the full gas exhibits a broad, asymmetric tail extending to high velocities ($\gtrsim 500\,\mathrm{km\,s^{-1}}$), the [C II]-bright component is significantly narrower and approximately Gaussian (section~\ref{subsec:kinematicstruct}), centered slightly below systemic ($v_{\rm rad} \sim -50\,\mathrm{km\,s^{-1}}$). This indicates that [C II] emission is dominated by dynamically colder gas associated with inflows, turbulence, and fountain recycling, rather than the fastest outflowing material. 

\item \textit{[C II] emitters ubiquitously exhibit outflowing motion but remain net inflow dominated.}  
Outflow rates rise with $L_{\rm [CII]}$ in all models; however, the net mass flux at the effective radius of [C II] emission remains negative across most luminosities and redshifts (section~\ref{subsec:mdotout}). This indicates that galaxy growth is primarily sustained by gravitationally driven accretion, with [C II] tracing the self-regulated baryon cycle rather than fast escaping winds.
 
\item \textit{[C II]-derived wind properties systematically misrepresent multiphase outflows.}  
Intrinsic mass-loading factors measured directly in the simulations decline with stellar mass for all gas phases. In contrast, mass-loading factors inferred from [C II] spectra using observational scaling relations increasingly underestimate the intrinsic multiphase $\eta_{\rm gas}$ toward low stellar masses, where the difference is largest (section~\ref{subsec:vout},~\ref{subsec:feedbackenergetics}). This discrepancy arises because [C II] emission preferentially traces the cold, low-velocity component of the gas flow and does not sample the high-velocity, hot phase that dominates the kinetic energy budget of galactic winds. In addition, spectroscopic velocity estimates derived from [C II] are weighted toward slower gas and therefore do not capture the full velocity distribution of the outflow, leading to systematically lower estimates of $\eta_{\rm gas}$ and wind energetics.

\item \textit{Broad [C II] components reflect both feedback-driven and gravitationally induced motions.}  
While stacked spectra reveal a broad component in the {\tt bursty-sn} model, the largest FWHM values are found in the {\tt smooth-sn} and {\tt hyper-sn} models, which exhibit weaker or no clear outflow signatures (section~\ref{subsec:broadoutflow}). Instead, FWHM correlates strongly with the circular velocity of the halo, indicating that line widths are primarily set by the depth of the gravitational potential rather than feedback driven winds alone.

\item \textit{Mode of feedback regulates [C II] kinematics and disk emergence.}  
$V_{\mathrm{rot}}/\sigma_0$ and disk fractions increase with decreasing redshift in all models (section~\ref{subsec:morphokinematic}), but the transition to disk-like systems ($V_{\mathrm{rot}}/\sigma_0 > \sqrt{3.36}$; see \citealt{Schreiber_2018}) occurs earliest in {\tt smooth-sn}, followed by {\tt hyper-sn}, and only by $z\sim5$ in {\tt bursty-sn}. By $z\sim5$, {\tt smooth-sn} and {\tt hyper-sn} reach disk fractions of $\sim40$--50\% at $M_* > 10^8$~M$_\odot$, compared to $\lesssim30\%$ in {\tt bursty-sn}. At lower stellar masses, disk fractions remain modest in all models, although $V_{\mathrm{rot}}/\sigma_0 > 1$ indicates ordered rotation even in systems that are not formally disk-like.

\item \textit{Kinematic classifications depend on the method used to identify disks.}  
Morpho-kinematic and rotation-curve-based classifications yield broadly similar disk fractions, but differ systematically in the inferred timing of disk formation by $\Delta z \sim 0.4$-1.2 (section~\ref{subsec:morphokinematic}). Morpho-kinematic methods identify disks earlier, while rotation-curve diagnostics impose a stricter requirement on the timing of disk formation.

\item \textit{[C II] kinematics provide an indirect probe of feedback at $z>5$.}  
The feedback prescriptions produce distinct trends in $V_{\mathrm{rot}}/\sigma_0$, FWHM, and the timing of disk-like kinematics, with {\tt smooth-sn} and {\tt hyper-sn} exhibiting earlier rotational support and larger line widths than {\tt bursty-sn}. These differences are directly reflected in the [C II] velocity structure and demonstrate that the kinematic properties of the cold gas depend sensitively on the mode of feedback.

\end{enumerate}

Overall, the {\tt SPICE} simulations indicate that the choice of supernova feedback model i.e {\tt smooth-sn}, {\tt bursty-sn}, or {\tt hyper-sn} sets the neutral gas content, the extent of [C II] emission, the energetics of the winds, and the timing of disk assembly in high-redshift galaxies. The observed diversity of [C II] luminosities, sizes, outflow signatures, and kinematic states can therefore arise from variations in feedback duty cycle at a fixed halo mass. Interpreting [C II] observations in the JWST era requires accounting for this multi-phase and time-variable nature of stellar feedback.

\section*{Acknowledgements}
 AB acknowledges support from the ERC synergy grant 101166930-RECAP. AB would like thank the cake bet with R{\"u}diger Pakmor which significantly sped up the work carried out for this paper. AB thanks Jack Birkin for helpful discussions. This work extensively used publicly software packages:  {\tt matplotlib} \citep{hunter2007matplotlib}, {\tt numpy} \citep{van2011numpy}, {\tt scipy} \citep{scipy}, {\tt astropy} and {\tt specutils} \citep{specutils}. A.B extends a thank you to the community of developers and those maintaining all of these packages. Computations for this work were performed on the HPC systems Freya and Orion at the Max Planck Computing and Data Facility (MPCDF).
\section*{Data Availability}
Data products employed in this work can be shared on reasonable request. 

\bibliographystyle{mnras}
\bibliography{biblio} 




\appendix
\section{Scaling relations from SPICE}
In this appendix we provide scaling relations for various [C II] derived quantities. Since we note a general lack of strong redshift evolution for most presented quantities, we provide fits taking all galaxies for all feedback models at $z=10-5$. Tables A1-A5 present fits for the $L_{\rm [C II]} -\rm SFR$,  $R_{\rm [C II]} - L_{\rm[CII]}$, $\dot{M}_{\rm out} - L_{\rm[CII]}$, and $v_{\rm out} - L_{\rm[CII]}$ relations for cold and all gas outflow velocities derived from the simulations, and finally $\eta_{\rm [C II]} - L_{\rm[CII]}$ relations derived from the simulations. We employ power law or second-order polynomials to produce fits, and choose the function that gives the lower chi-squared errors. Additionally, all fits are performed for $L_{\rm [CII]} > 10^4 \rm L_\odot$.

\label{sec:appendixA}
\begin{table}
\centering
\caption{Power-law fits for the $L_{\rm [C II]}$ - SFR relation for three feedback models. All galaxies in the range $z=10-5$ are included when calculating the best fit.}
\label{tab:lcII_sfr_fit}
\begin{tabular}{|p{3cm}||p{5cm}|}
\hline
Model & Scaling Relation \\
\hline
\hline
{\tt bursty-sn} & $L_{\rm [C II]} = 10^{6.74} \times \mathrm{SFR}^{1.00} - 300$ \\
{\tt smooth-sn} & $L_{\rm [C II]} = 10^{7.31} \times  \mathrm{SFR}^{1.04} + 100$ \\
{\tt hyper-sn}  & $L_{\rm [C II]} = 10^{7.20} \times  \mathrm{SFR}^{1.17} + 10$ \\
\hline
\end{tabular}
\end{table}

\begin{table}
\centering
\caption{Second-order polynomial fits between $R_{\rm [C\,II]}$ and  $L_{\rm [C II]}$ for the three feedback models and all galaxies in the range $z=10-5$.}
\label{tab:rcii_lcii_polyfit}
\begin{tabular}{|p{3cm}||p{5cm}|}
\hline
Model & Scaling Relation \\
\hline
\hline
{\tt bursty-sn} & $R_{\rm [C\,II]} = -10^{-16.17} \, L_{\rm [C\,II]}^2 + 10^{-7.64} \, L_{\rm [C\,II]} + 0.459$ \\
{\tt smooth-sn} & $R_{\rm [C\,II]} = -10^{-18.22} \, L_{\rm [C\,II]}^2 + 10^{-8.65} \, L_{\rm [C\,II]} + 0.480$ \\
{\tt hyper-sn} & $R_{\rm [C\,II]} = -10^{-18.01} \, L_{\rm [C\,II]}^2 + 10^{-8.46} \, L_{\rm [C\,II]} + 0.577$ \\
\hline
\end{tabular}
\end{table}

\begin{table}
\centering
\caption{Fitting functions for outflow rates as a function of $L_{\rm [C II]}$ estimated at effective radii of [C II] emission for the three feedback models and all galaxies in the range $z=10-5$.}
\label{tab:outflow_scaling}
\begin{tabular}{|p{3cm}||p{5cm}|}
 \hline
 \multicolumn{2}{|c|}{}\\
 Model & Scaling Relation \\
 \hline
 \hline
 {\tt bursty-sn} & $\dot{M}_{\rm out} = 10^{-4.61} \times L_{\rm [C II]}^{0.74} + 0.01$\\
 {\tt smooth-sn} & $\dot{M}_{\rm out} = 10^{-1.68} \times L_{\rm [C II]}^{0.27} + 0.58$ \\
 {\tt hyper-sn} & $\dot{M}_{\rm out} = 10^{0.63} \times L_{\rm [C II]}^{0.07} - 8.81$  \\
 \hline
\end{tabular}
\end{table}

\begin{table}
\centering
\caption{Second-order polynomial fits for outflow velocities as a function of $\log_{10}(L_{\rm [C\,II]})$ for the three feedback models and all galaxies in the range $z=10-5$.}
\label{tab:vout_lcii_polyfit}
\begin{tabular}{|p{3cm}||p{5cm}|}
\hline
Model & Scaling Relation \\
\hline
\hline
\multicolumn{2}{|c|}{$v_{\rm out, [C\,II]}$}\\
\hline
{\tt bursty-sn} & $v_{\rm out} = 4.53\times (\log_{10} L_{\rm [C\,II]})^2 - 38.77\times \log_{10} L_{\rm [C\,II]} + 102.13$ \\
{\tt smooth-sn} & $v_{\rm out} = 2.21\times, (\log_{10} L_{\rm [C\,II]})^2 - 20.86\times \log_{10} L_{\rm [C\,II]} + 66.45$ \\
{\tt hyper-sn}  & $v_{\rm out} = 1.56\times, (\log_{10} L_{\rm [C\,II]})^2 - 10.34\times \log_{10} L_{\rm [C\,II]} + 40.84$ \\
\hline
\hline
\multicolumn{2}{|c|}{$v_{\rm out, gas}$}\\
\hline
{\tt bursty-sn} & $v_{\rm out} = 26.67\times (\log_{10} L_{\rm [C\,II]})^2 - 212.71\times \log_{10} L_{\rm [C\,II]} + 493.11$ \\
{\tt smooth-sn} & $v_{\rm out} = 2.93\times (\log_{10} L_{\rm [C\,II]})^2 - 13.01\times \log_{10} L_{\rm [C\,II]} + 126.58$ \\
{\tt hyper-sn}  & $v_{\rm out} = -18.43\times (\log_{10} L_{\rm [C\,II]})^2 + 298.55\times \log_{10} L_{\rm [C\,II]} - 863.18$ \\
\hline
\end{tabular}
\end{table}

\begin{table}
\centering
\caption{Scaling relations between $\log_{10}(\eta_{\rm _{\rm gas}})$ and $\log_{10}(L_{\rm [C II]})$ for the three feedback models between $z=10-5.$}
\label{tab:eta_lcii_fit}
\begin{tabular}{|p{3cm}||p{5cm}|}
\hline
Model & Scaling Relation \\
\hline
\hline
{\tt bursty-sn} & $\log(\eta_{\rm gas}) = 0.261\times  \log(L_{\rm [C II]}) - 6.484$ \\
{\tt smooth-sn} & $\log(\eta_{\rm gas}) = 0.079\times  \log(L_{\rm [C II]}) - 35.260$ \\
{\tt hyper-sn}  & $\log(\eta_{\rm gas}) = -0.030\times  \log(L_{\rm [C II]})^2 - 0.068 \times \log(L_{\rm [C II]}) + 3.429$ \\
\hline
\end{tabular}
\end{table}

\section{Impact of inclination angle on $R_{\rm eff}$}
\label{sec:AppendixB}
\begin{figure}
    \centering
\includegraphics[width=0.99\linewidth]{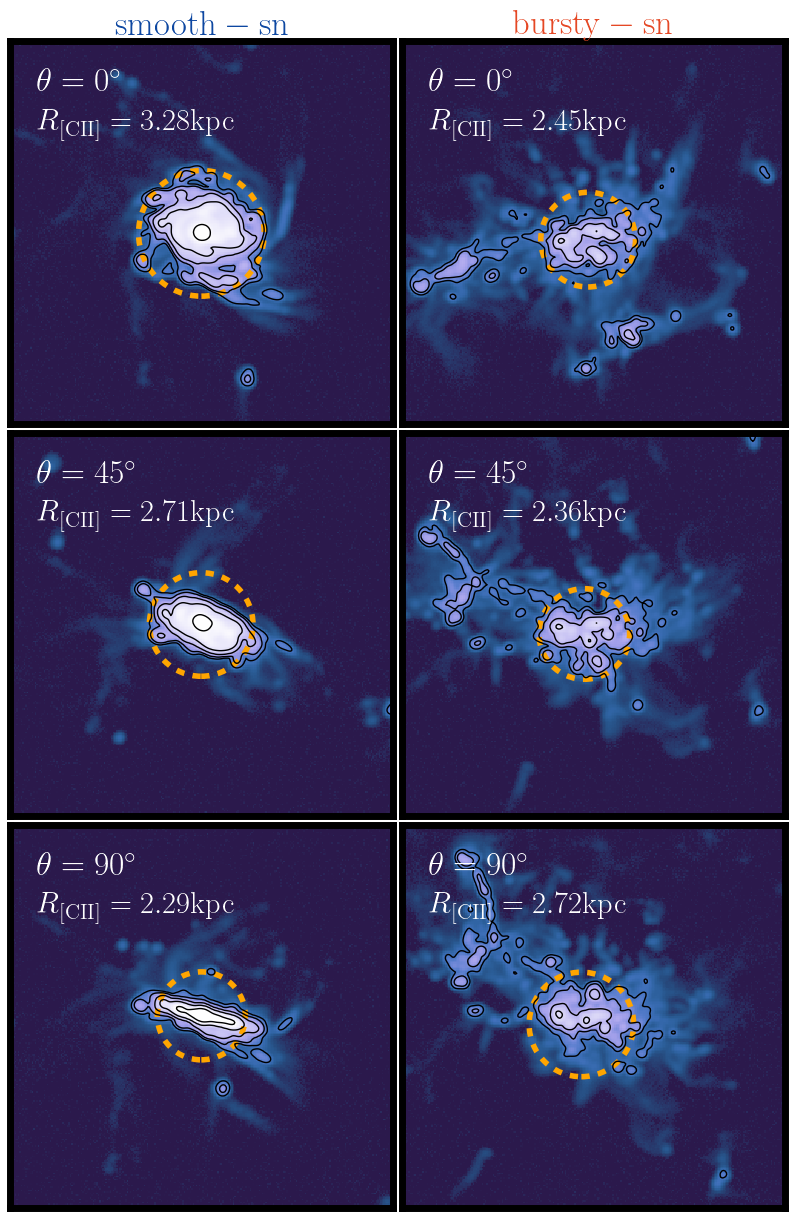}
    \caption{Surface brightness maps for the most massive halo in {\tt smooth-sn} (left column) and {\tt bursty-sn} (right column) at $z=5$ at three inclinations: $0^\circ$, $45^\circ$ and $90^\circ$. $R_{\rm [C II]}$ (as calculated in section~\ref{sec:SyntheticdataCII}) is represented visually with an orange circle and number indicates the size for the respective inclination.}
    \label{fig:Orientation}
\end{figure}

Figure~\ref{fig:Orientation} illustrates the impact of inclination angle on the inferred [C II] morphology and size for the {\tt smooth-sn} and {\tt bursty-sn} models. While the {\tt smooth-sn} galaxy exhibits a coherent, disk-like structure across all orientations, the {\tt bursty-sn} system appears more irregular and clumpy, with extended asymmetric features. The inferred [C II] radius, $R_{\rm [CII]}$, varies strongly with inclination for {\tt smooth-sn}, decreasing for more edge-on views, consistent with projection effects of a flattened disk, whereas {\tt bursty-sn} shows minor systematic variation due to its disturbed and clumpy morphology. These results demonstrate that both the apparent structure and size of [C II] emission are sensitive to viewing angle, with stronger orientation dependence in systems with ordered disk-like gas distributions.

\bsp	
\label{lastpage}
\end{document}